\documentclass{article}

\usepackage{fullpage}
\usepackage{tikz}
\usepackage[all]{xy}
\usepackage{color}
\usepackage{float}
\usepackage{graphicx}
\usepackage{amsmath}
\usepackage{wrapfig}
\usepackage{epstopdf}
\usepackage{amssymb}
\usepackage{authblk}

\definecolor{Sion}{rgb}{.45,0.05,.85}

\definecolor{forest}{rgb}{.01,.6,.2}

\definecolor{wine}{rgb}{.6,.01,.3}

\title{Knowledge gaps in the early growth of semantic networks}

\author[1]{Ann E. Sizemore}
\author[2]{Elisabeth A. Karuza}
\author[1]{Chad Giusti}
\author[1,3,*]{Danielle S. Bassett}

\affil[1]{Department of Bioengineering, University of Pennsylvania, Philadelphia, PA 19104 USA}
\affil[2]{Department of Psychology, University of Pennsylvania, Philadelphia, PA 19104 USA}
\affil[3]{Department of Electrical and Systems Engineering, University of Pennsylvania, PA 19104 USA}

\begin{document}

\maketitle

\begin{abstract}
Understanding the features of and mechanisms behind language learning can provide insights into the general process of knowledge acquisition. Recent methods from network science applied to language learning have advanced the field, particularly by noting associations between densely connected words and acquisition. However, the importance of sparse areas of the network, or knowledge gaps, remains unexplored. Here we create a semantic feature network in which words correspond to nodes and in which connections correspond to semantic similarity. We develop a new analytical approach built on principles of applied topology to query the prevalence of knowledge gaps, which we propose manifest as cavities within the network. We detect topological cavities of multiple dimensions in the growing semantic feature network of children ages 16 to 30 months. The pattern of cavity appearance matches that of a constrained null model, created by predefining the affinity of each node for connections. Furthermore, when word acquisition time is computed from children of mothers with differing levels of education, we find that despite variation at the word level, the global organization as measured by persistent homology remains comparable. We show that topological properties of a node correlate with filling in cavities better than simple lexical properties such as the length and frequency of the corresponding word. Finally, we show that the large-scale architecture of the semantic feature network is topologically accommodating to many node orders. We discuss the importance of topology in language learning, and we speculate that the formation and filling of knowledge gaps may be a robust feature of knowledge acquisition.
\end{abstract}

\section*{Introduction}

Formal analysis of the mechanisms driving knowledge acquisition remains a foundational area of research in cognitive science. In the domain of word learning, behavioral evidence suggests that this process is mediated in part by various properties of words at an individual level, such as the frequency of a given word or the extent to which it evokes a mental image \cite{duff2012role,ambridge2015ubiquity}. Recently, tools from network science have offered a means of examining how lexical acquisition might also be mediated by higher-order relationships between many words, or the network topology underlying input that is available to the learner \cite{karuza2016local}. Under this approach, words are typically represented by the nodes of the network, while shared semantic or sound-based associations can be used to construct edges between them \cite{hills2009longitudinal, goldstein2014influence, steyvers2005large}. Broadly, evidence indicates that learners are particularly sensitive to how densely connected a given word is relative to others words in a network.

Because these previous studies focus on the areas of the network that have been learned by children, they have left open the question, ``How do those words not yet known affect learning?'' More precisely, as children produce new words in the semantic network, does their attained semantic network contain any knowledge gaps, or voids where a word is missing? Since edges correspond to meaning, a gap in the network suggests a unifying concept that is not yet understood. In the network science formalism, such knowledge gaps in a growing network correspond to topological cavities that are born, and then later filled in with the addition of new nodes and edges. We propose that characterizing the evolution of these cavities in a semantic network offers unique insight into lexical organization in children, and we investigate whether knowledge gaps might serve as a useful proxy for the difficulty associated with acquiring feature-based concepts.

To answer these questions, we employ concepts and tools from applied topology that allow us to detect topological cavities within a growing semantic network. The specific network that we study is a semantic network in which words are given a weight (and ordering) derived from the month that the word was produced by toddlers aged 16-30 months. Though such node-weighted networks are commonly observed in biology, they are challenging to analyze because common tools from network science, such as traditional graph metrics, can account for weighted edges, but not for weighted nodes. To address this challenge, we develop a formalism and construction that transforms any node-weighted system into a sequence of binary graphs analyzable by both topological data analysis and standard tools of network science including graph metrics. Specifically, we transform the growing semantic feature network into a sequence of binary graphs called a \emph{filtration}, with one new node (corresponding to one new word) added at each step. We call this sequence of graphs built from a node-ordered network a \emph{node-filtered order complex} (hereafter referred to as the n-order complex for brevity), inspired by the order complex defined for edge-weighted networks \cite{giusti2015clique}. We can then compute \emph{persistent homology} \cite{carlsson2009topology,zomorodian2005computing}, which tracks the formation and possible filling in of topological cavities of different dimensions throughout a filtration.

By encoding the growing semantic feature network of children ages 16-30 months as a n-order complex, we use persistent homology to ask if topological cavities -- corresponding to knowledge gaps -- form and then fill in throughout the learning process. In contrast to expectations derived from a growing network null model, we find a collection of long-persisting cavities of varying dimensions. Interestingly, the pattern of cavity formation suggests that the semantic network is organized under heavy constraints. We adjudicate between conflicting hypotheses that topological cavities might either be robust to or vary systematically with the nature of input available to the learner, in this case indexed by the mother's level of education. We observe at most minor differences in topological cavity existence despite random variation in node order; on average, any ordering of words produces a similar topological signature. Our results suggest that these topological cavities might be a conserved feature of the learning process, and that semantic network growth is a robust phenomenon that can accommodate many local changes without abrupt restructuring of its large-scale organization.

\section*{Materials and Methods}

\subsection*{Growing semantic network construction}
We constructed a 120-node semantic feature network with node ordering following the procedure outlined in \cite{hills2009longitudinal}. Specifically, we used the \textit{Wordbankr} package \cite{frank2016wordbank} which contains data from the MacArthur-Bates Communicative Development Inventory (MB-CDI) \cite{dale1996lexical}. This database contains which of 541 English words 5511 toddlers ages 16-30 months could produce, as recorded via parental report. For each word we calculate the month at which $\geq 50\%$ of children could produce the word \cite{hills2009longitudinal}. Within one month, words are ordered according to the percentage of children producing each word, resulting in a total ordering of words.

To form a node-ordered network we represent words as nodes and connect two nodes (words) if they share a semantic feature within the McRae feature list \cite{mcrae2005semantic}. These semantic features were derived from adult norming data from 725 adults and are organized into categories based on feature type. When an individual generates features for a given concept, these features are interpretations of the abstract concept that are constructed for the sole purpose of description \cite{barsalou2003abstraction}. Then feature norms offer a unique understanding of representation, which varies across and within individuals, and which often results in a collection of distinguishing features (as opposed to general features of many words) \cite{mcrae2005semantic}, making them useful for modeling and theory testing \cite{hampton1979polymorphous,wu2009perceptual,devlin1998category,moss2002emergence}. We use all feature categories excluding encyclopedic and taxonomic, which are unlikely to be accessible to toddlers \cite{hills2009longitudinal}. All words included in both the McRae and Wordbankr databases were used in our final semantic network.

\subsection*{Persistent homology}
Below we include a brief description of persistent homology. We refer the interested reader to \cite{carlsson2009topology,zomorodian2005computing,ghrist2008barcodes} and the Appendix for more details.

Given a graph $G = (V,E)$, we begin by assigning a $k$-simplex to each set of $(k+1)$ all-to-all connected nodes, called a $(k+1)$-clique. Recall a $k$-simplex is the convex hull of $k+1$ affinely positioned points, and the clique complex $X(G)$ is the collection of all simplices defined by appropriately sized cliques of $G$. Within a clique complex, a $k$-cycle is a closed path of $k$-simplices. Note that a $k$-cycle either encloses a collection of higher dimensional simplices (called a $k$-boundary) or a topological cavity. We call two $k$-cycles equivalent if their set difference is a $k$-boundary. This creates an equivalence relation, and the number of non-trivial equivalence classes indicates the number of topological cavities of dimension $k$. As is common in the literature, we refer to an equivalence class of $k$-cycles as a $k$-cycle.

We construct a graph filtration, or sequence of graphs each included in the next, from the growing binary network by letting $G_n$ be the graph after the addition of node $n$, with $G_n = G_{n-1} \cup \{n\}$ and all of its connections between node $n$ and nodes $1,\dots, n-1$. This process produces a graph filtration indexed by nodes, and consequentially a node-indexed filtration of clique complexes. We call this filtration of clique complexes induced by the binary graph and node-ordering the \emph{node-filtered order complex}. This filtration of clique complexes allows us to map the $k$-cycles in $X(G_n)$ to $k$-cycles in $X(G_{n+1})$. We call these cycles -- which are tracked throughout the filtration -- \emph{persistent cycles}. We compute the persistent homology \cite{carlsson2009topology,zomorodian2005computing} in dimensions 1--3 using the Eirene software \cite{henselmanghrist16}.

\subsection*{Models of n-order complexes}
For each model n-order complex, we generate 1000 instances and provide MATLAB code and detailed descriptions at the Filtered Network Model Reference (filterednetworkmodelref.weebly.com).

\subsection*{Measures for correlation calculations}
For correlations with the number of persistent cycles killed at each node, we compute the following graph statistics using \cite{rubinov2010complex} on the binary semantic feature network: node degree, clustering coefficient, and betweenness centrality. The degree of a node is the number of edges incident to the node. The clustering coefficient measures the connectivity of a node's neighbors, calculated by the ratio of existing triangles to the number of triangles possible. Precisely,
\begin{equation}
C_n = \frac{2t_n}{k_n(k_n -1)}
\end{equation}
where $t_n$ is the number of triangles formed by node $n$ and its neighbors \cite{watts1998collective}.

Additionally, we inquire whether the centrality or number of shortest paths passing through a node might correlate with the number of cycles killed. In particular, we define the betweenness centrality \cite{kintali2008betweenness} of a node as
\begin{equation}
BC_n = \sum_{s,t,n, s\neq t\neq n} \frac{\lambda_{n}(s,t)}{\lambda(s,t)}
\end{equation}
with $\lambda(s,t)$ being the number of shortest paths between nodes $s$, $t$, and $\lambda_n(s,t)$ being the number of such paths passing through node $n$.

\section*{Results}

\begin{figure}[h]
	\centering
	\includegraphics[width = .7\textwidth]{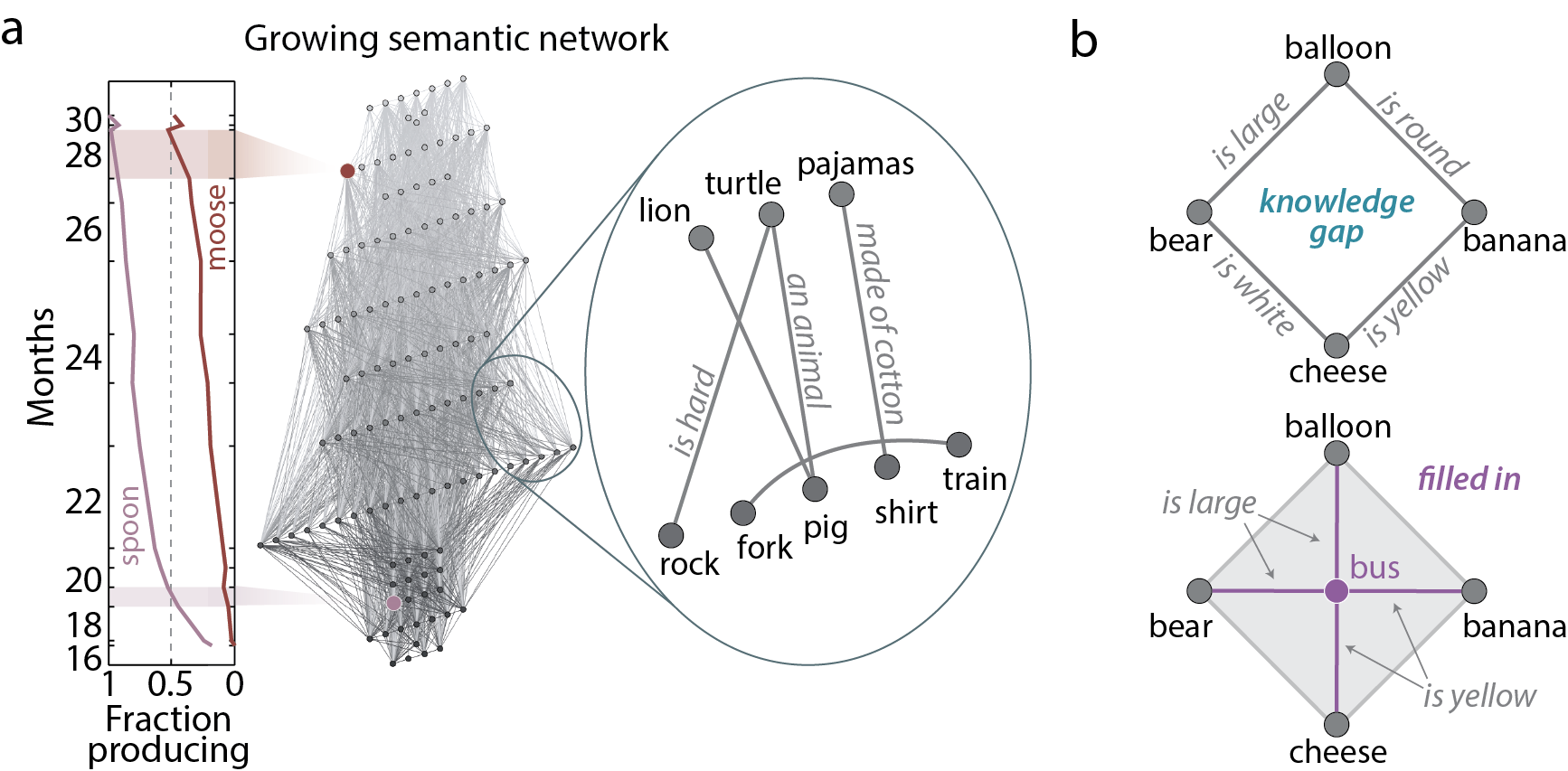}
	\caption{\textbf{Knowledge gaps manifesting as topological cavities within the growing semantic network.} \emph{(a)} (Left) Word ordering is given based on the month at which $50\%$ of children produce each word. As an example, the word `spoon' is first produced by $\geq50\%$ of children at 19 months, so it is placed at the appropriate location within the growing complex (purple node, towards the bottom). The word `moose' is similarly placed at the 28 month mark (sienna node, towards the top). (Right) Semantic features connect nouns (corresponding to nodes), forming the semantic network. (Center) Combining the binary feature network and word production times creates a growing semantic network with nodes entering based on the first month at which $\geq 50\%$ of children can produce the word. \emph{(b)} A `knowledge gap' could be seen as a topological void within the semantic network. The connection pattern between `balloon', `bear', `cheese', and `banana' leave a gap within the graph (top), but the addition of the node corresponding to `bus' and its connections fills in the cavity (bottom).}
	\label{fig:1}
\end{figure}

To begin, we construct an ordering on 120 nouns derived from the first month at which $\geq$50\% of children between the ages of 16 and 30 months can produce each word (Fig.~\ref{fig:1}a, left) \cite{frank2016wordbank}. Multiple words could be first produced within one month, so we create a total ordering by sorting words learned within one month by ascending percentage of children producing each word. We next form a binary semantic feature network with 120 nouns as nodes, and with edges connecting words that share a property or function (Fig.~\ref{fig:1}a, right) \cite{mcrae2005semantic}. Together, the word ordering and binary network pair assemble into the growing semantic network \cite{hills2009longitudinal}, with the node added at step $n$ connecting to all of its neighbors added at steps $1,\dots, n-1$ (Fig.~\ref{fig:1}a, middle).

We are interested in the presence of knowledge gaps, which we hypothesize manifest as voids within the growing semantic network that exist only for a finite number of months. For example, in Fig.~\ref{fig:1}b (top), the words `balloon', `bear', `cheese', and `banana' connect in a pattern that leaves a hole within the network. If the word `bus' is learned later, this word connects to each of `balloon', `bear', `cheese', and `banana', so that there is no longer a void within the network. When a void in the network is extinguished, we say the knowledge gap is \emph{filled in} (Fig.~\ref{fig:1}b, bottom). The features of interest in such a network are then the nodes responsible for filling in the cavity, which correspond to the temporarily missing words.

\subsection*{Detecting cavities in node-ordered networks}

To identify topological cavities within the growing semantic network, we use a method from applied topology called persistent homology, which returns the (1) number, (2) dimension, and most importantly (3) longevity of topological cavities within a growing network, all of which we will more rigorously define in this section.

 \begin{figure*}[t]
 	\centering
 	\includegraphics[width = .8\textwidth]{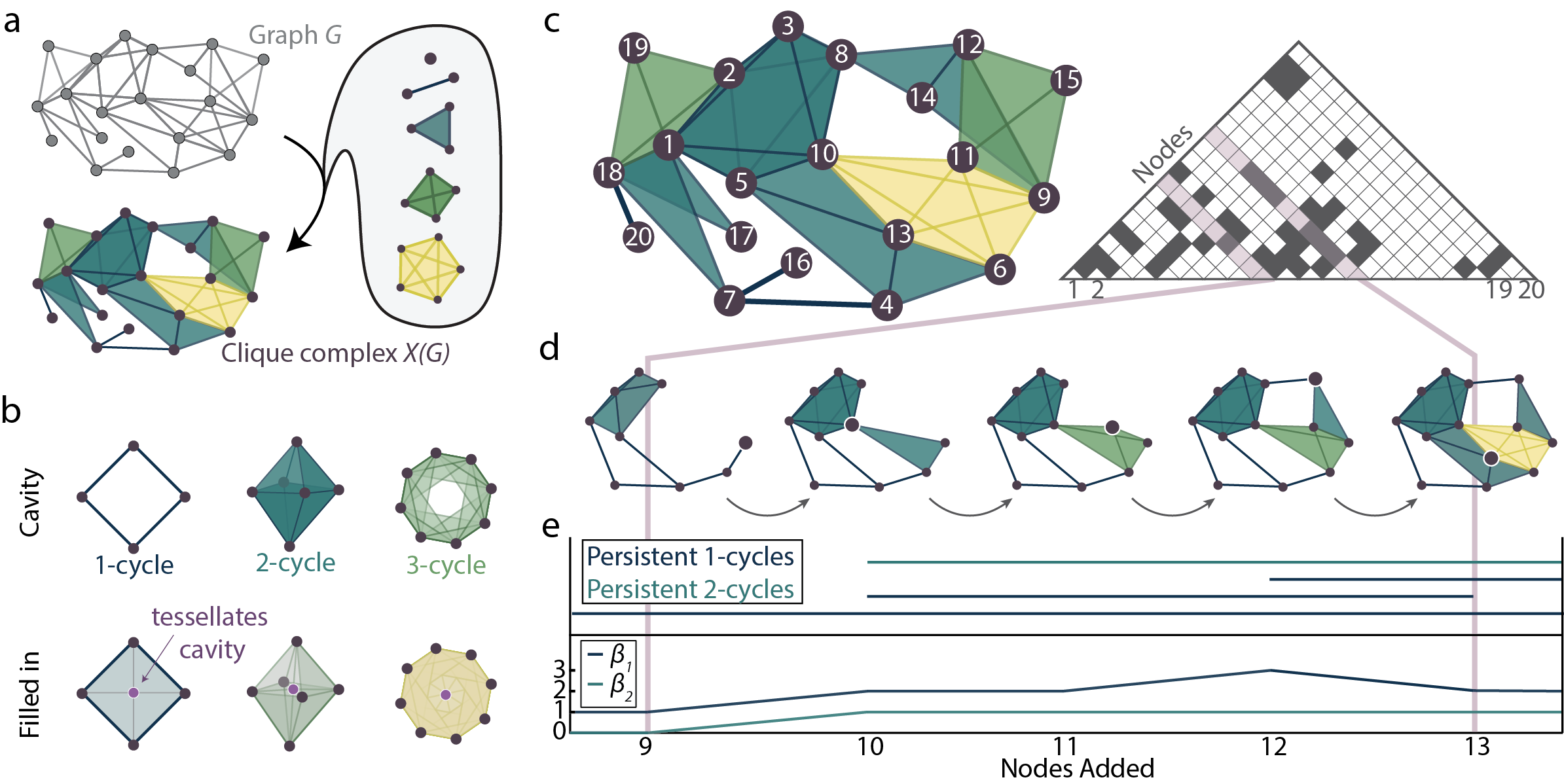}
 	\caption{\textbf{Persistent homology detects longevity of topological cavities within node-filtered order complexes.} \emph{(a)} Example graph $G$ (top) and its clique complex $X(G)$ (bottom) created by filling in cliques, or all-to-all connected subgraphs of $G$. \emph{(b)} Examples in dimensions 1-3 of cavities enclosed by cycles (closed paths of cliques) (top) and how an added node can tessellate a cycle thus filling in the cavity (bottom). \emph{(c)} The clique complex from \emph{(a)} with an ordering on the nodes (left), and the associated ordered adjacency matrix (right). \emph{(d)} Steps 9-13 in the filtration created by taking the node-filtered order complex of the clique complex $X(G)$ in \emph{(c)} and the shown ordering. At each step a new node is added along with its connections to nodes already present in the complex. \emph{(e)} Barcode (top) and Betti curves (bottom) for the example node-filtered order complex. The barcode shows the lifespan of a persistent cavity as a bar extending from $[birth,$ $death)$ node, and the Betti curves count the number of $k$-dimensional cavities as a function of nodes added. Lavender lines through \emph{(c)}, \emph{(d)}, and \emph{(e)} connect the adjacency matrix row $i$ to the clique complex at step $i$ and to the persistent homology outputs. }
 	\label{fig:2}
 \end{figure*}
 
Before we discuss growing graphs, we outline the process of detecting cavities in a single binary network. Given a binary graph $G$, we first translate our graph into a combinatorial object on which we can perform the later computations. Instead of a simple graph described by nodes and edges, we allow all groups of completely connected nodes to define entities. Formally, we create the \emph{clique complex} $X(G)$, a collection of all the \emph{cliques}, or all-to-all connected subgraphs, in the network. In Fig.~\ref{fig:2}a, we depict this process as `coloring in' the graph $G$ to build the clique complex $X(G)$. For example, we color in 1-cliques (nodes), 2-cliques (edges), 3-cliques (triangles), and so on, giving us higher dimensional information about the structure\footnote{Precisely we assign a $k$-simplex to each $(k+1)$-clique within $G$ to create the clique complex. See Appendix for definitions and details.}.

Now with our graph encoded as a clique complex, we can use \emph{homology} to detect cavity-surrounding motifs of edges, triangles, tetrahedra, and higher dimensional analogs (Fig.~\ref{fig:2}b). Loops of edges form 1-cycles, loops of triangles form 2-cycles, and loops of tetrahedra form 3-cycles. For example, Fig.~\ref{fig:2}b shows cavity-surrounding cycles of each dimension on the top row, while those on the bottom are tessellated by higher-dimensional cliques formed with the purple node. Homology distinguishes between cavity-surrounding loops and those tessellated by higher cliques, thereby returning detailed information about the mesoscale architecture of the complex. In particular, homology detects \emph{equivalence classes} of $k$-cycles, with two $k$-cycles being in the same equivalence class if their symmetric difference is a collection of higher dimensional cliques (see Appendix for details). By abuse it is common to refer to an equivalence class of $k$-cycles as a $k$-cycle, and we will adopt this abbreviated description throughout the remainder of the paper. To summarize: homology counts the number of cavities in each dimension of a clique complex constructed from a binary graph.

While this approach is hypothetically useful, our data describes a \emph{growing} network instead of a single binary graph, so we cannot simply compute its homology as above described. However, notice that we get a binary graph after the addition of each new node, and that the binary graph $G_n$ created after the addition of node $n$ is a subgraph of $G_{n+1}$ for all $n$. This sequence of objects (here, graphs) with $G_n \subseteq G_{n+1}$ is called a \emph{filtration} (Fig.~\ref{fig:sfig_filts}, top and Fig.~\ref{fig:sfig_ph2}b, top). If we construct a filtration of binary graphs $G_n$, we immediately gain a filtration of clique complexes $X(G_n)$ with $X(G_n)\subseteq X(G_{n+1})$ necessarily true since $G_n \subseteq G_{n+1}$ (Fig.~\ref{fig:sfig_ph2}b, middle). For example, using the ordering and clique complex in Fig.~\ref{fig:2}c, Fig.~\ref{fig:2}d illustrates the described filtration of clique complexes for steps 9-13 (addition of nodes 9-13), with new nodes (outlined in white) connecting to any neighbor already in the complex. We call the filtration of clique complexes created from a growing network the \emph{node-filtered order complex} (see Appendix for further details), inspired by the order complex in \cite{giusti2015clique}. The order complex creates a filtration of clique complexes from an edge-weighted network using the edge weights to induce an edge ordering. Here, the node filtered order complex (which we shorten to n-order complex for brevity) can be completely defined by the pair $(G,s)$ with $G$ a binary graph and $s$ the ordering of vertices, possibly induced by a weighting on the nodes.

Finally, at each node addition we can map the clique complex $X(G_n)$ into the next $X(G_{n+1})$, so we can follow cycles, and consequentially cavities, throughout the filtration. For example, the addition of node 10 creates a cavity surrounded by a 1-cycle, which persists in the complexes $X(G_{10}), X(G_{11}), X(G_{12})$ until it is tessellated with the addition of node 13. The barcode plot in the top of Fig.~\ref{fig:2}e records this persistent cavity as a horizontal line running throughout the duration of this persistent cavity, or its \emph{lifetime}. We call the node at which the cavity begins the \emph{birth} node, and we call the node at which the cavity is tessellated the \emph{death} node. Thus, the lifetime is formally $death-birth$. The number of cavities of dimension $k$ at each step (node addition) in the filtration is recorded in the Betti curves $\beta_k(n)$ and shown in the bottom of Fig.~\ref{fig:2}e for the n-order complex of Fig.~\ref{fig:2}d. Tracking these persistent cycles throughout a filtration is called \emph{persistent homology} \cite{carlsson2009topology,zomorodian2005computing}, which -- in summary -- allows us to extract the number and dimension along with the longevity of topological cavities throughout the growth process.

\subsection*{Topology of generative growing network models}

Before approaching knowledge gaps in the semantic network, we first pause to ask if and how persistent homology can distinguish randomness from structure within artificial models of growing graphs (equivalently n-order complexes), which will additionally help us to gain an intuition for processes creating particular cavity existence patterns. Specifically, we test four growing graph models with varying degrees of predefined structure. Each model constructs a binary graph by assigning a probability to the existence of each edge as a function of one or both parent nodes. We call these \emph{generative} models because they construct \textit{de novo} both the binary graph and the node order. For the following models we let the node ordering $s$ simply be $1:N$ with $N$ the total number of nodes.

\begin{figure}[h]
	\centering
	\includegraphics[width = 3in]{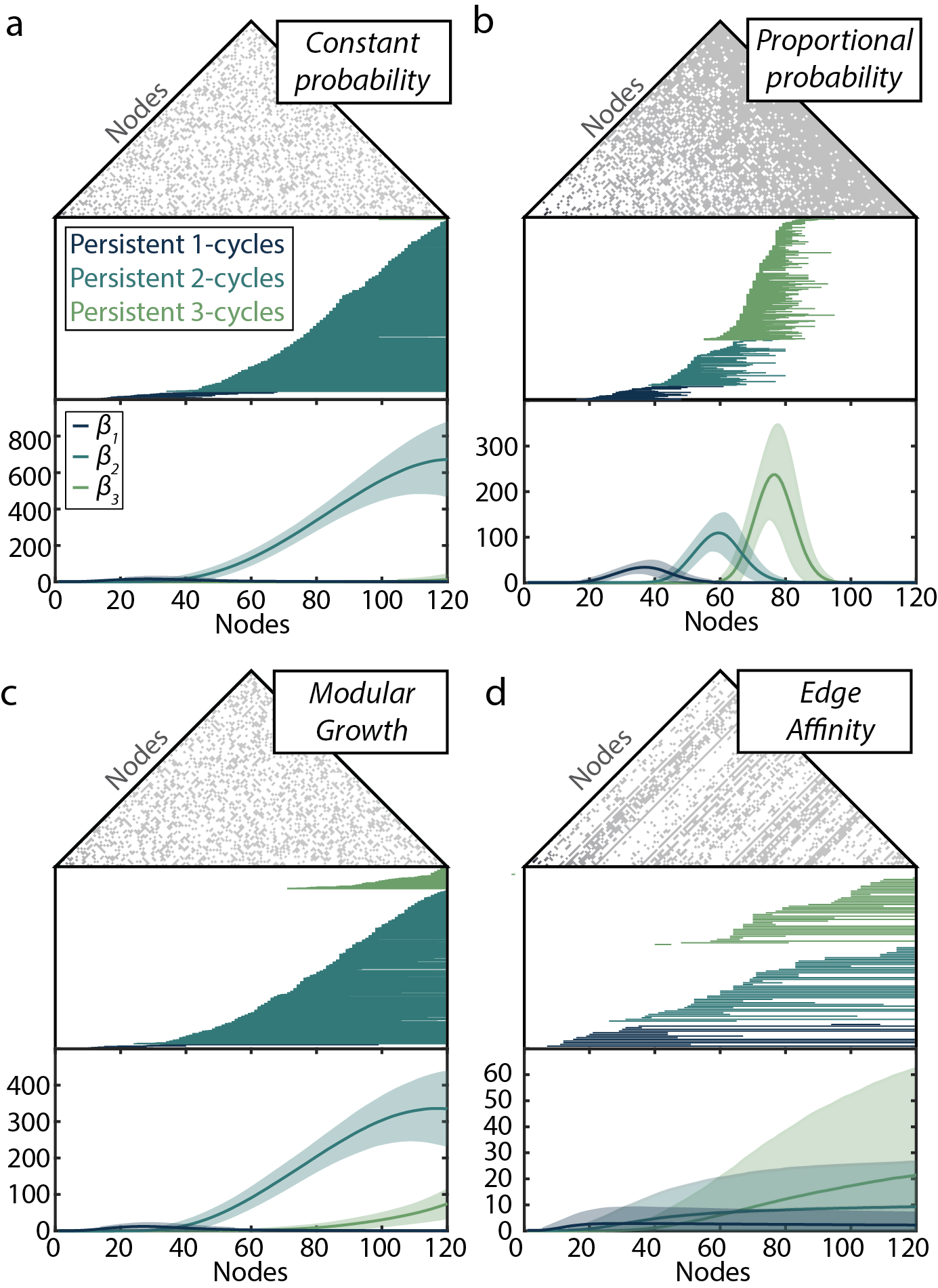}
	\caption{\textbf{Persistent homology distinguishes random from structured generative models of node-filtered order complexes.} Representative adjacency matrix (top), associated barcode plot (middle), and average Betti curves (bottom) for the \emph{(a)} constant probability, \emph{(b)} proportional probability, \emph{(c)} modular growth, and \emph{(d)} edge affinity models. Shaded regions in Betti curve plots indicate $\pm 2$ standard deviations.}
	\label{fig:3}
\end{figure}

The most basic (and random) model that we tested assigns each edge entering the graph with the addition of node $n$ a probability $p(n) = c \in [0,1]$ of existing. We call this model the \emph{constant probability} model, and we show its persistent homology in Fig.~\ref{fig:3}a with $p(n) = 0.3$. The next model, the \emph{proportional probability} model, attaches edges from node $n$ to all nodes $1, \dots, n-1$ with probability $p(n) = n/N$ (Fig.~\ref{fig:3}b). Next, we generate a modular network composed of four equal-sized communities, and we refer to it as the \emph{modular growth} model (Fig.~\ref{fig:3}c). We randomly assign nodes to communities; edges added with node $n$ exist with probability $p_{in}$ between $n$ and nodes within its community, and with probability $p_{out}$ between $n$ and members of different communities. In our final generative model, each node $n$ is assigned an affinity $a_n$ for edges such that when node $n$ is added, edges between node $n$ and $m = 1, \dots, n-1$ exist with probability $p(n,m) = \frac{a_m}{\max(\vec{a})}$ with $\vec{a} = (a_1, a_2, \dots, a_N)$ being the vector of affinities. We call this model the \emph{edge affinity} model, and we show results for this model with affinities given by a random permutation of $(1:120)^2$ in Fig.~\ref{fig:3}d. For each of these models, we choose parameters so as to produce graphs whose edge density closely matches the edge density of the empirically measured semantic network, $\sim 0.3$.

The constant probability model generates growing graphs with the least amount of imposed structure, producing hundreds of persistent 2-cycles that never die. The $\beta_2$ curve on average dominates the Betti curves and we observe few if any persistent 1-cycles or 3-cycles. The proportional probability model instead shows an increasing trend of Betti curve peaks with increasing dimension. Additionally, all persistent cycles of dimensions 1-3 die by around node 100, as later nodes are likely to tessellate cavities. Interestingly, the modular growth model produces Betti curves dominated by dimension 2, similar to the constant probability model. Though this model produces networks with high modularity throughout the node addition process, we see qualitatively similar properties between this and the constant probability model. Still, the density of within-community connections restricts the maximum height of the $\beta_2$ peak and drives the creation of persistent 3-cycles.

In contrast to these null models, we expect the growing semantic structure to be organized according to external constraints: there exist (external) properties of nodes that do not fluctuate based on the current state of the network. If a node (word) has some aptitude for connections (similar to many other words), such an aptitude should not change as the network grows. Such a constancy is unlike that observed in, for example, a preferential attachment process but is explicitly accounted for in our edge affinity model. Interestingly, we observe far fewer persistent cycles of each dimension in the edge affinity model than in the previous models. Furthermore, we observe a pattern of increasing peaks of persistent cycles as we move to higher dimensions. These results demonstrate that a growing process constrained by the external constraint of constant edge affinity will yield fewer topological cavities than that of a more random growth process.

\subsection*{Gaps in the growing semantic feature network}

\begin{figure}
	\centering
	\includegraphics[width = 6in]{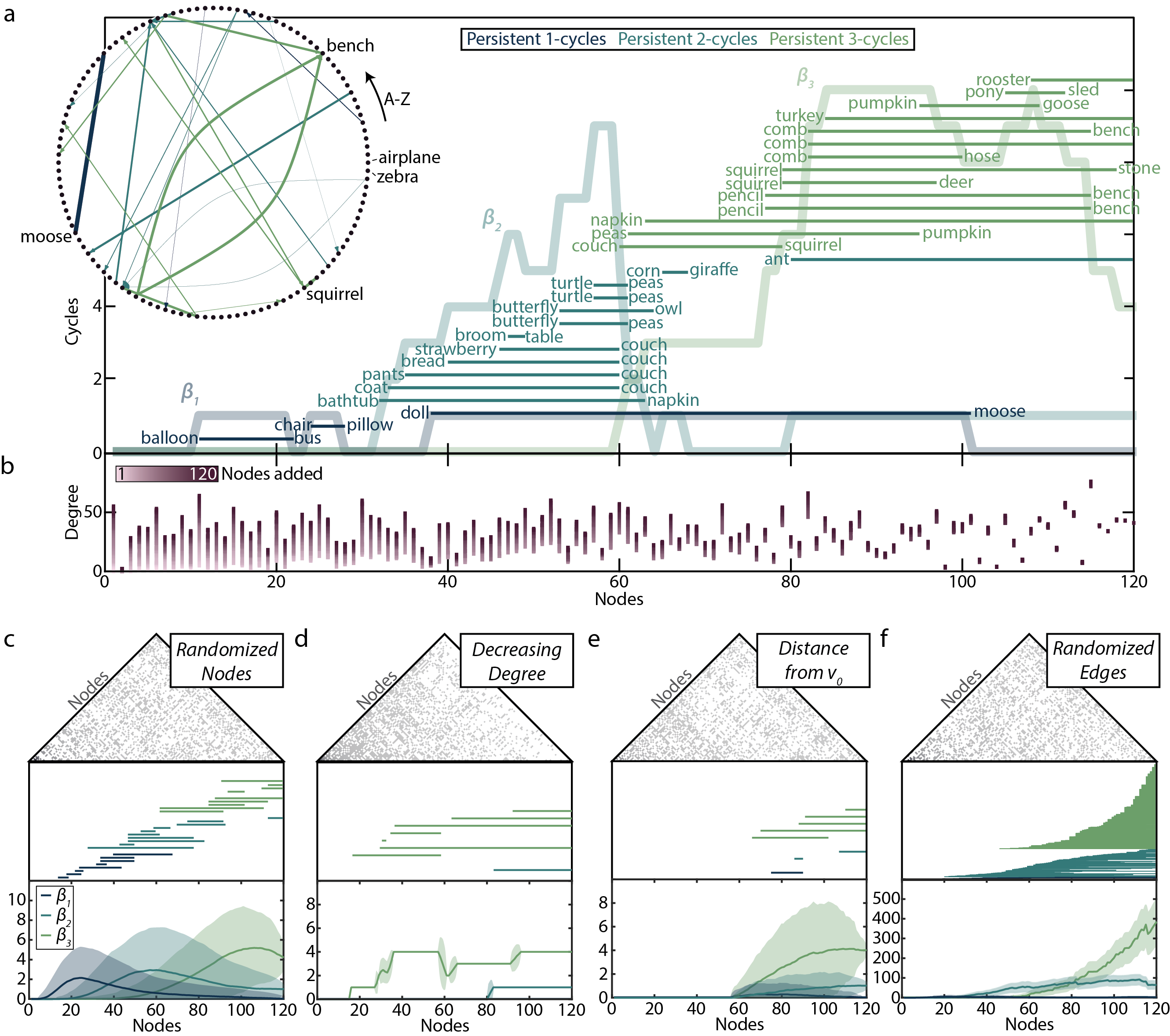}
	\caption{\textbf{Topological cavities form and die within the semantic network with a pattern that is resistant to random node reordering.} \emph{(a)} Barcode and Betti curves for the growing semantic network. The word added when the cavity is born (killed) is written on the left (right) of the corresponding bar. (Inset) Graph of persistent cycles with words as nodes in alphabetical order. An edge for each persistent cavity in \emph{(a)} exists from the birth to the death node. Edges are weighted by the persistent cycle lifetime and colored according to the dimension. \emph{(b)} The degree of each node throughout the growth process. Color indicates the number of nodes added. Representative adjacency matrix (top), associated barcode (middle), and average Betti curves (bottom) for the \emph{(c)} randomized nodes, \emph{(d)} decreasing degree, \emph{(e)} distance from $v_0$, and \emph{(f)} randomized edges models. Shaded areas of Betti curves indicate $\pm2$ standard deviations.}
	\label{fig:4}
\end{figure}

Now that we have developed some intuition for the structure detected by persistent homology, we ask if topological cavities exist within the growing semantic network and if so, what information such cavities might provide about the learning process. We observe multiple persistent cavities of dimensions 1-3, most of which die before the 30 month mark (Fig.~\ref{fig:4}a). The Betti curves show increasing peaks with larger dimensions as more nodes are added. Next, we ask which nodes (words) enter the growing graph when persistent cycles are born or killed. We list these words next to the corresponding bar, and we show persistent cycle birth and death nodes visually in the inset of Fig.~\ref{fig:4}a as a persistent cycle network, with words ordered alphabetically and an edge for each persistent cavity emanating from the birth node and terminating at the death node. The edge thickness is proportional to the corresponding cavity lifetime (index of death minus index of birth) and the edge color indicates cavity dimension. We observe nodes generally begin or kill one or no persistent cycles, with a few exceptions such as `bench', `peas', and `couch'. Persistent cavities that never die (have a death time of $\inf$) are not shown. The fact that very few cavities exist in the final semantic network suggests that knowledge gaps not only form, but must also evolve and are extinguished during the learning process.

Next, we ask if there are simple rules by which cavities form and evolve in the growing semantic network. We notice that nodes added late in the growth process have higher chances of having a high degree at the time of their addition than nodes added early in the growth process. Thus, one might hypothesize that the empirically observed pattern of Betti curves follows simply from a pattern of higher-connectivity nodes added throughout the filtration. Contrary to this simplistic expectation, we observe instead that the degree of nodes varies considerably across time with no salient trend of either a decreasing or increasing node degree (Fig.~\ref{fig:4}b). Indeed, when the final node is added there exists great variability in node degrees when plotted in the order of node addition. This complexity motivates a more thorough effort to model the growth process to infer underlying mechanisms of cavity formation and evolution, which we turn to next.

The persistent homology of a growing network is classically used to infer global organizational properties, and we can therefore use this tool to understand the growing semantic network from a global perspective. We begin by comparing the persistent homology of the growing semantic network to the generative models of Fig.~\ref{fig:3}. We observe that the edge affinity model generates n-order complexes with the most similar persistent homology to that of the growing semantic network. Upon closer inspection, we observe that the largest difference between the Betti curves of the affinity model and growing semantic network stems from the likelihood that cavities die. All but five persistent cavities in the growing semantic network die by node 120, while in the edge affinity model the barcodes show the majority of persistent cycles never die. Moreover the comparisons with models shown in Fig.~\ref{fig:3} strongly imply that the evolving architecture of the growing semantic network is highly non-random, as the Betti curve peaks are not even of the same magnitude between the growing semantic and random n-order complex models. Taken together, these results suggest that the growing semantic network topology during learning is non-random and that node constraints such as a fixed affinity for connections might play a role in the evolving architecture.

To further probe mechanisms guiding the evolution of the growing semantic structure, we construct \emph{derived} n-order complex models that begin with the semantic network and alter either the node ordering or edge placement to determine which (if either) explains the observed evolving architecture. Beginning with the influence of node order, we compute the persistent homology of the \emph{randomized nodes} n-order complex model, which retains the binary graph of semantic feature connections but randomly permutes the node order. We observe strikingly similar persistent homology between this model (Fig.~\ref{fig:4}c) and the growing semantic network, though note that any n-order complex built with the same binary graph $G$ will necessarily have the same homology at $G_N$, which limits the variability in persistent homology. Next we keep the binary network of semantic connections but now order the nodes by decreasing degree (Fig.~\ref{fig:4}d) or by their topological distance from the first node (Fig.~\ref{fig:4}e), which we call the \emph{decreasing degree} and \emph{distance from $v_0$} models, respectively (any ties are randomly permuted). Both of these more engineered models exhibit persistent homology that is less similar to the growing semantic network than the randomized nodes model, suggesting that neither learning the most connected words first nor learning those with the shortest semantic distance to the first word can account for the evolution of the growing semantic network. Finally, we keep the node ordering constant but now randomly rewire edges while preserving node degree; we call this the \emph{randomized edges} model and note that it is similar in spirit to the configuration model \cite{bender1978asymptotic,maslov2002specificity}. We observe a highly random persistent homology signature as described by the Betti curves and barcodes (Fig.~\ref{fig:4}f), suggesting that the pattern of semantic feature connections between words is more important in explaining knowledge gap formation than the order in which those words are produced by children.

\subsection*{Influence of or robustness to maternal education level}

\begin{figure}[t]
	\centering
	\vspace{-10pt}
	\includegraphics[width = 3in]{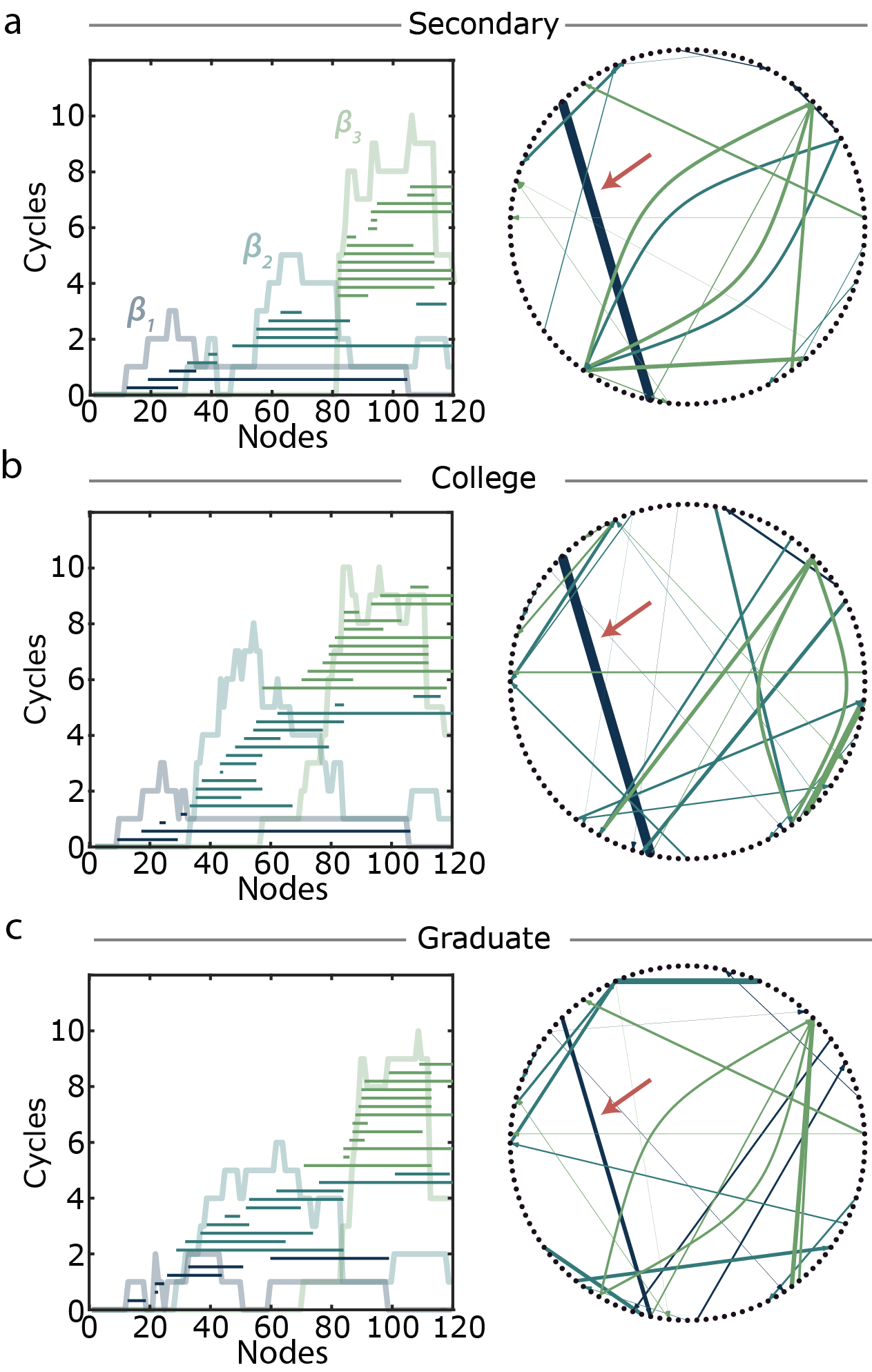}
	\caption{\textbf{Global semantic network architecture is consistent across maternal education levels despite local variations.} (Left) Betti curves with barcodes overlaid and (right) persistent cycle networks for the \emph{(a)} \textit{secondary}, \emph{(b)} \textit{college}, and \emph{(c)} \textit{graduate} growing semantic networks. Red arrow in persistent cycle networks indicates a persistent cavity born and killed by the same word in each of the three education levels.}
	\label{fig:5}
	\vspace{-10pt}
	
\end{figure}

In the previous sections, we demonstrated that knowledge gaps not only exist but are created and filled in throughout early semantic learning. Knowledge gaps might correspond to learning relatively difficult concepts, and one could hypothesize that gaps would occur more frequently in the growing semantic networks of children with mothers having achieved a higher level of education, which -- along with socioeconomic status \cite{hoff2005socioeconomic,schwab2016language} -- can significantly impact child-directed speech and learning \cite{dollaghan1999maternal}. Yet, a contrary hypothesis is that the laws of evolution predispose the growing semantic networks in children to be relatively robust to variations in the environment. To adjudicate between these two conflicting hypotheses, we create three distinct growing semantic networks from children with mothers whose highest education was some or all of secondary school, some or all of college, and some or all of graduate school. Then we have the same binary network for each of the three networks, but the ordering of nodes has now changed. We label these networks the \textit{secondary}, \textit{college}, and \textit{graduate} growing networks, respectively. We compute the persistent homology and find no trend of increasing topological cavity number or lifetimes (Fig.~\ref{fig:5}a-c), despite the differences in word ordering.

Since at a global level the persistent homology of the three growing networks varies little, we next ask if the same words correspond to nodes killing persistent cavities in each growing network. Figure \ref{fig:5} (also see Fig.~\ref{fig:sfig_education1}) shows persistent cycle networks for each of the three growing networks with nodes ordered and placed alphabetically as in Fig.~\ref{fig:4}a. This visualization allows for comparison of nodes that begin and kill persistent cavities across the three growing networks. For example, a persistent 1-cycle is seen beginning at `doll' and ending at `pony' in each of the \textit{secondary}, \textit{college}, and \textit{graduate} growing networks (indicated by the red arrow). We observe that while a few node pairs begin and end persistent cycles in each of the \textit{secondary}, \textit{college}, and \textit{graduate} growing networks, generally node pairs do not begin and end persistent cycles, or at least persistent cycles of the same dimension, across each of the education levels.

\subsection*{Characterizing the manner in which knowledge gaps are extinguished}

\begin{figure}[h]
	\centering
	\includegraphics[width = .9\textwidth]{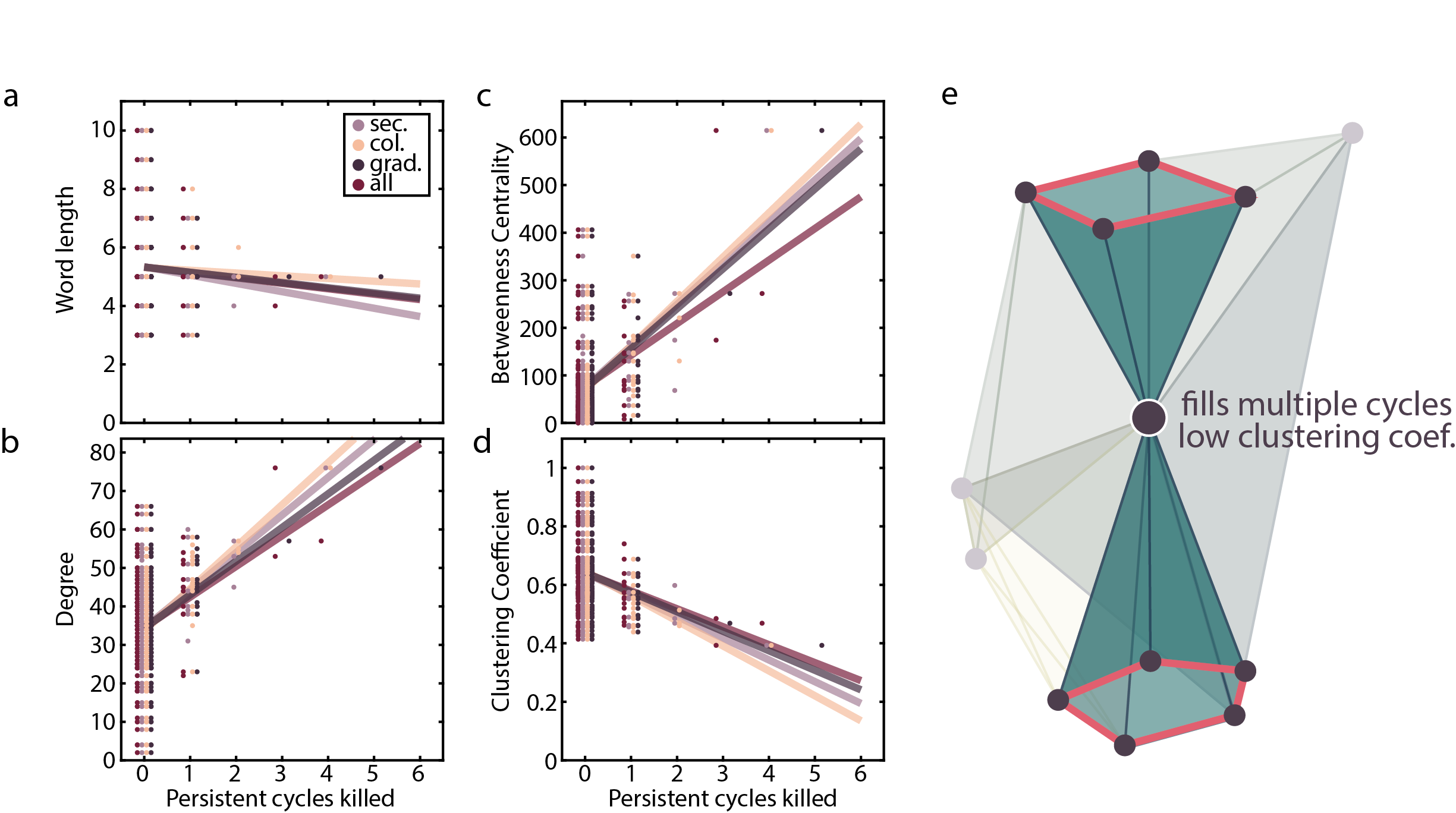}
	\caption{\textbf{Number of persistent cycles killed correlates with topological properties instead of lexical features.} Scatter plots of the number of persistent cycles killed by each node against \emph{(a)} corresponding word length, \emph{(b)} node degree, \emph{(c)} betweenness centrality, and \emph{(d)} clustering coefficient. Lines of best fit overlaid. \emph{(e)} Example node (outlined in white) that kills multiple cavities while retaining a low clustering coefficient. Triangles formed by the cavity-killed node highlighted, and cycles tessellated outlined in red.}
	\label{fig:6}
\end{figure}

In the previous sections, we have shown that knowledge gaps are created and later filled in with similar rates despite differences in maternal education. These observations motivate our final effort to determine if particular properties of the nodes or their corresponding words increase the likelihood of a node tessellating cavities. For each of the \textit{secondary}, \textit{college}, \textit{graduate}, and original all-inclusive growing semantic network barcodes, we count the number of persistent cavities killed by each node. Since these cavity-killing nodes correspond to temporarily missing words, one might hypothesize that these corresponding words are more difficult to learn. We use a simple proxy for word difficulty: word length. However we find no significant correlation between the number of cycles killed and word length (Fig.~\ref{fig:6}a, $p>0.1$ for all growing networks, Pearson correlation coefficient). Additionally, we ask if the frequency with which caregivers use words when speaking to children could play a role in cavity filling, expecting lower-frequency words to be more difficult for children to learn. However, again we observe no significant correlation between frequency and number of persistent cycles killed (Fig.~\ref{fig:app_corr}). These results suggest that simple word descriptors such as length and frequency do not predict a word's tendency to fill in knowledge gaps.

Next we test if topological characteristics of nodes -- rather than their non-topological statistics such as length and frequency -- might better explain the number of persistent cycles killed. To address this question, we study the number of persistent cycles killed against node degree, centrality, and clustering coefficient (Fig.~\ref{fig:6}b-d). While node degree and betweenness centrality are positively correlated with the number of persistent cycles killed ($p<0.01$) as expected, the clustering coefficient shows a negative correlation ($p<0.01$). Initially, this result might appear counterintuitive because to cone (fill in) a cycle, a node must by definition create many triangles. Yet, if we combine this result with the positive correlations of persistent cycle killing to node degree and betweenness, we can construct a toy example of a possible cavity-killing node neighborhood. Shown in Fig.~\ref{fig:6}e, the central node outlined in white tessellates two cycles when added (cycles and coning triangles highlighted), but also has a low clustering coefficient. Taken together, we suggest that the connectivity pattern of words within the semantic network better predicts the tendency of that word to fill a knowledge gap than simple lexical features of the words themselves.

\section*{Discussion}

In this study, we query the existence of knowledge gaps manifesting as topological cavities within the growing semantic feature network of toddlers. Using persistent homology and the formalism of node-filtered order complexes, we find that such knowledge gaps both form and are often later filled in throughout the learning process. We observe that the global architecture of the growing semantic network is similar to that of a constrained generative model. Furthermore we report similar persistent homology across growing semantic networks of children from mothers with differing education, and we find that this pattern of topological cavity existence remains present after node order randomization, but not after edge rewiring. Together these results suggest that knowledge gaps are robust features of word production order and that the global topology of the semantic feature network is resilient to local alterations induced by node reordering.

\subsection*{The accommodating topology of the growing semantic network}

Understanding the growing semantic network through the lens of persistent homology offers a novel perspective on the nature of the learning process. Previous research has provided evidence supporting the influence of network topology on many types of language networks including those constructed from phonological \cite{arbesman2010structure,siew2013community} and syntactic relations \cite{corominas2009ontogeny,vcech2009word}. Here we observe that the persistent homology of the growing semantic network follows a regular pattern throughout the majority of the learning process, indicating an organized and potentially predictable growth pattern.

Yet when we consider the node (word) level, the fine-scale topology (node degree) varies considerably and does not suggest a predictable addition pattern. Furthermore if we permute the order of the nodes uniformly at random, we recover similar global topology to that observed in the unpermuted network. We describe the semantic network topology as accommodating, since its large-scale architecture changes little despite variations in small-scale inputs (new nodes with differing degrees). Previous studies show that the order of word learning in children depends on multiple variables including word frequency \cite{brent2001role,huttenlocher1991early}, parental interaction \cite{hart1995meaningful}, and communication quality \cite{hirsh2015contribution}. We speculate that part of the learning mechanism might include a global accommodating topology that develops similarly in children despite variations in input (order of words learned) due to differing environments.

Though many possibilities for word production order yield similar persistent homology, we observe that the global structure disintegrates if we randomly rewire the network edges while preserving the degree of each node (Fig.~\ref{fig:4}). This finding -- together with the results described in the previous paragraph -- suggests that the higher order connectivity patterns between words, instead of individual word properties such as time of production, have a greater impact on the evolving global structure of the semantic network. This observation raises the important question of whether this resilience to node reordering and emphasis on fixing relations between words is restricted to the English language, or whether these phenomena are general properties of language networks. Previous research points to the similarity of word networks across languages \cite{youn2016universal} and supports the hypothesis that similar global patterns would be observed in languages other than English. Yet, the differences in node (word) connectivity patterns within this structure could offer insight into subtle distinguishing features between different languages \cite{goddard2008cross}.

\subsection*{Hints toward novel learning mechanisms}

The presence of persistent topological cavities of multiple dimensions in the growing semantic network offers novel insights into the learning process. One might expect that when one grows one's vocabulary, one tends to learn words that are similar to words already known. Such a process would leave few if any knowledge gaps, corresponding to topological cavities, within the network, and should be well-modeled by a topological-distance-from-initial-node rule. By contrast, we observe the salient presence of topological cavities in a growing semantic network that is best modeled by an edge-affinity rule. Yet, it is also important to acknowledge that the recovered gaps are not simply gaps in the final semantic network itself. Instead by the age of 30 months all but five gaps have been filled in by other words. Since we observed cavities in the growing semantic network irrespective of the mother's education, we speculate that knowledge gaps that form \emph{and} die may themselves be a feature of the semantic learning process. If indeed these knowledge gaps represent learning a more difficult word or concept, and filling in the created gap with intermediate concepts as they are later added, then these cavities may be a natural part of the learning process. As an extension to more explicit learning in a classroom, one could ask if cavities exist as students learn other subjects as well, in particular math and science where reaching for an understanding of distant or difficult concepts may create higher numbers of cavities or longer-lived persistent cavities.

Many models exist for semantic networks in which edges are defined by word association such as can be estimated from a free association task or other metrics. Yet, models for the growing semantic \emph{feature} networks as we study here are scarce if present, and efforts to use preferential attachment or close variations have not been successful in capturing the feature network's development \cite{hills2009longitudinal}. The difference between the previously defined preferential attachment model and the affinity model that we introduce here is that the likelihood of new connections for each node evolves based on how the network has already grown in the former, while the latter only relies on predefined properties of nodes. As the semantic network a child is able to produce grows, the child is likely already sensitive to semantic relationships in their external world even before they acquire the label attached to a previously unnamed object. This sensitivity might explain why the affinity model better captures the topological properties of the growing semantic network.

One possible way that the above concept can manifest in our encoding is that labels for features are allowed to exist in the network before a child is able to produce this label. For example, if a child produces `cheese' and `bus' but not `yellow' at a given time, the lack of the child's ability to produce `yellow' does not mean that `cheese' and `bus' are not still both yellow and thus are connected. Similarly, since each word exists and connects to other words in the external world, a word will always have the same affinity for others. For example, any new animal with legs will always be ready to connect to all other objects that have legs, regardless of which of these words is known to the child. Furthermore, we speculate that words corresponding to nodes with high affinity may generally be \emph{polysemous} words, or those with multiple meanings, thus increasing the likelihood of connections to other words within the semantic network \cite{sole2015ambiguity}. Overall our results are consistent with an externally constrained topology of the semantic network, as suggested in \cite{hills2009longitudinal,steyvers2005large}.

\subsection*{Topology in growing processes}

Growing networks are implemented in multiple systems including contagion propagation \cite{taylor2015topological}, distribution networks in biological systems \cite{papadopoulos2016embedding}, and social networks \cite{jin2001structure}. Consequentially, numerous methods exist for their analysis \cite{holme2012temporal}. For example, representing a growing process as a dynamic network or directed graph (or directed dynamic graph) would allow for analyses with those corresponding sets of tools \cite{chowdhury2016persistent,sizemore2017dynamic}. Though persistent homology for node-weighted systems is not a novel concept \cite{taylor2015topological,hofer2017deep,courtney2017weighted}, we suggest the formalism presented here as a practical mode of encoding such systems, in which both graph metrics and topological data analysis can be simultaneously applied. Additionally, these models hearken to previous theoretical studies including node-exchangeable graphs \cite{aldous1985exchangeability,hoover1979relations}, growing simplicial complexes \cite{courtney2017weighted,bianconi2016emergent}, and random clique complexes \cite{kahle2013limit}.

Furthermore, we propose that the persistent homology of a n-order complex at the level of persistent features (as opposed to the more commonly studied global structure) may be more easily interpretable than that of an edge-weighted network. In particular, here individual words initiate and terminate persistent cycles. In many biological contexts, nodes are objects with attached empirical observations and metadata. The analyses of such systems might include analyzing the number of persistent cycles a node begins or kills in relation to this metadata\footnote{though assigning full responsibility of persistent cycles to individual cliques of any size should always be done with care \cite{bendich2015stabilizing}}. Additionally the n-order complex is invariant under any monotonic (rank-preserving) transformation of the node weights which is a noted benefit for applications to noisy experimental data \cite{giusti2015clique,petri2013topological}. Topics suitable for the node-filtered complex encoding and subsequent analyses include tracking information dissemination through brain networks \cite{mivsic2015cooperative}, signaling cascades in protein interaction networks \cite{vinayagam2011directed}, sound propagation on force chains \cite{bassett2012influence}, contagion spreading on social networks, and information transfer throughout enzyme architecture after allosteric effector binding \cite{cockrell2013new}. Broadly, the formality presented in this study may be useful for ``filling in'' open questions from multiple areas of science.

\subsection*{Conclusions}

In conclusion, we offer a novel perspective on the growing semantic feature network of toddlers that highlights the persistence of knowledge gaps in contrast to the formation of densely connected clusters. Using the node-filtered order complex formalism and persistent homology, we reveal the existence of such knowledge gaps and their persistence as children age. Furthermore we provide evidence supporting the notion that the gaps in the network will exist despite differences in word production times, and we speculate that these gaps are an important and general component of the learning process.

\section*{Acknowledgments}
The authors thank Leonardo Torres, Dr. Tina Eliassi-Rad, and Brennan Klein for helpful discussions. This work was supported by the National Science Foundation CAREER PHY-1554488 to DSB. The authors also acknowledge support from the John D. and Catherine T. MacArthur Foundation, the Alfred P. Sloan Foundation, the Paul G. Allen Foundation, the Army Research Laboratory through contract number W911NF-10-2-0022, the Army Research Office through contract numbers W911NF-14-1-0679 and W911NF-16-1-0474, the National Institute of Health (2-R01-DC-009209-11, 1R01HD086888-01, R01-MH107235, R01-MH107703, R01MH109520, 1R01NS099348 and R21-M MH-106799), the Office of Naval Research, and the National Science Foundation (BCS-1441502, CAREER PHY-1554488, BCS-1631550, and CNS-1626008). The content is solely the responsibility of the authors and does not necessarily represent the official views of any of the funding agencies.

\bibliographystyle{plain}
\bibliography{bibfile}

\clearpage
\newpage

\onecolumn
\renewcommand{\thefigure}{S\arabic{figure}}

\setcounter{figure}{0}

\section*{Supplementary Information}

\subsection*{Details of topological methods}

This section outlines the details of encoding the growing semantic network as a filtration and computing the persistent homology. We devote significant real estate to the encoding and briefly describe persistent homology since it is more thoroughly discussed elsewhere \cite{carlsson2009topology,zomorodian2005computing,ghrist2008barcodes,hatcher2002algebraic}.

\subsubsection*{Node-weighted networks and induced filtrations}

Our motivation comes from data described most naturally as a network with weights on the nodes. Such systems can arise from protein-protein interaction networks with protein expression as node weights, structural brain networks with region activity as node weights, or a social network with time of contamination as nodes weights. Though generally classic graph statistics do not extend easily to networks with node weights, we present a construction that allows the simultaneous computation of both graph statistics and persistent homology on node-weighted networks.

It is important to note that we derive our inspiration from an object generated from edge-weighted networks called the \emph{order complex} \cite{giusti2015clique}. Given a graph with edge weights, we obtain an ordering on the edges by decreasing edge weights. Then we create a sequence of graphs, $G_0 \subset G_1 \subset \dots G_{|E|}$ with each $G_i$ the graph containing the $i$ highest ranked edges. This sequence of graphs is called the \emph{order complex} of the weighted network (or corresponding symmetric weighted matrix).

Now returning to node-weighted networks, we get a node ordering from the decreasing node vales. Then from this node ordering $s$ and graph $G$ with $N$ nodes we can similarly construct a sequence of graphs $G_0 \subset G_1 \subset \dots \subset G_N$ with $G_n$ containing the first $n$ nodes in $s$ and any connections between these nodes which exist in $G$. We call this sequence of binary graphs the \emph{node-filtered order complex} of a node-weighted network, denoted nord$(G,w)$ with $G$ the binary graph and $w:N \rightarrow \mathbb{R}$ the function assigning node weights. For brevity we often refer to this object as the n-order complex. Note then the n-order complex is completely determined by the pair $(G,w)$ or $(G,s)$ with $s$ the node ordering.

The n-order complex retains intrinsic developmental aspects of the node-weighted system, and furthermore allows for computation of both common binary graph metrics and persistent homology on these objects. Most graph metrics are not generalizable to include orders (weights) on the nodes, but if we instead construct the n-order complex, we now can compute such metrics on each $G_n$ in the filtration.\\

\noindent\emph{Construction of n-order complex into order complex}

In practice, the software that we use to compute the persistent homology expects oder complexes. Then to compute the persistent homology of a growing network, we encode the associated n-order complex as an order complex. We will show by construction that any n-order complex $\text{nord}((G,s))$ can be translated into a weighted network $M$ with each $G_i$ in the filtration of $\text{nord}((G,s))$ equal to $G'_i$ in the filtration of $\text{ord}(M)$.

Given a node-ordered network $(G,s)$, we construct the node-filtered order complex $\text{nord}((G,s))$. Then each $G_i$ is a binary graph and can be written as an $N \times N$ binary symmetric matrix $M_i$ (with $N-n$ extra padding rows and columns). Let $M = \sum_{i=0}^N M_i$. Now we have created a real-valued symmetric matrix $M$ that encodes a weighted network with the highest edge weights corresponding to the earliest added edges. If we then create $\text{ord}(M)$ with binary graphs $G_i'$, then by our construction it must be that $G'_i = G_i$ for all $i = 0, \dots, N$. Thus any n-order complex can be written as a weighted matrix $M$ with $\text{ord}(M) = \text{nord}((G,s))$. It is important to note that to get $G_i = G_i'$, we must assume that either all nodes always exist and we use the ``growing graph" concept to describe how edges are added, or for both we do not include nodes in the graph until they have a neighbor. Finally, if we denote by $\mathcal{N}$ the set of filtrations achievable by node-weighted networks and $\mathcal{O}$ the set of filtrations achievable by edge-weighted networks, by the above discussion we must have $\mathcal{N} \subseteq \mathcal{O}$.

\begin{figure}[h!]
	\centering
	\includegraphics[width=5in]{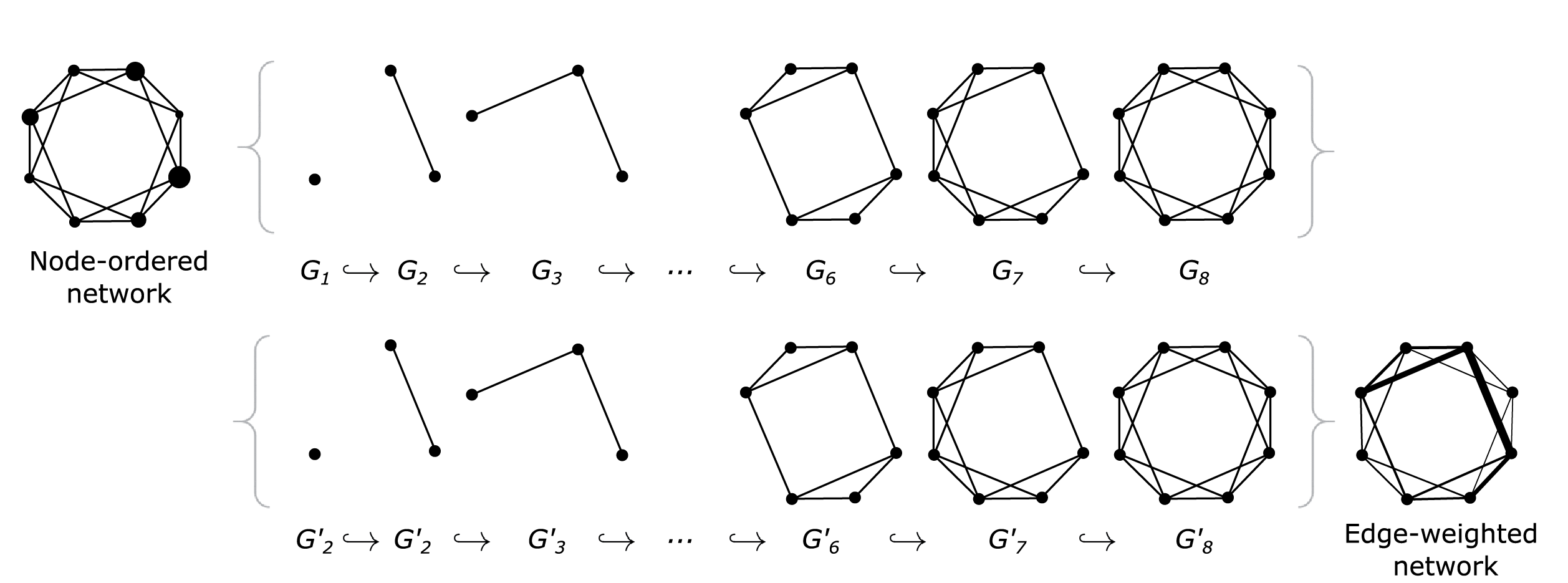}
	\caption{Filtration of a node-ordered network (top) and equivalent edge-weighted network yielding the same graph filtration (bottom).}
	\label{fig:sfig_filts}
	
\end{figure}

To examine the reverse relation, take the $3$-clique with nodes $a$, $b$, and $c$ with weights $e_{a,b} = 3$, $e_{b,c} = 2$, $e_{a,c} = 1$. The resulting filtration could not be created from a node-weighted network, since the final edge $e_{a,c}$ would be added to two nodes that already exist. In other words, it could not be added as the result of adding a new node. Therefore $\mathcal{N} \subsetneq \mathcal{O}$.

\subsubsection*{Persistent homology}

\begin{figure}
	\centering
	\includegraphics[width = 4in]{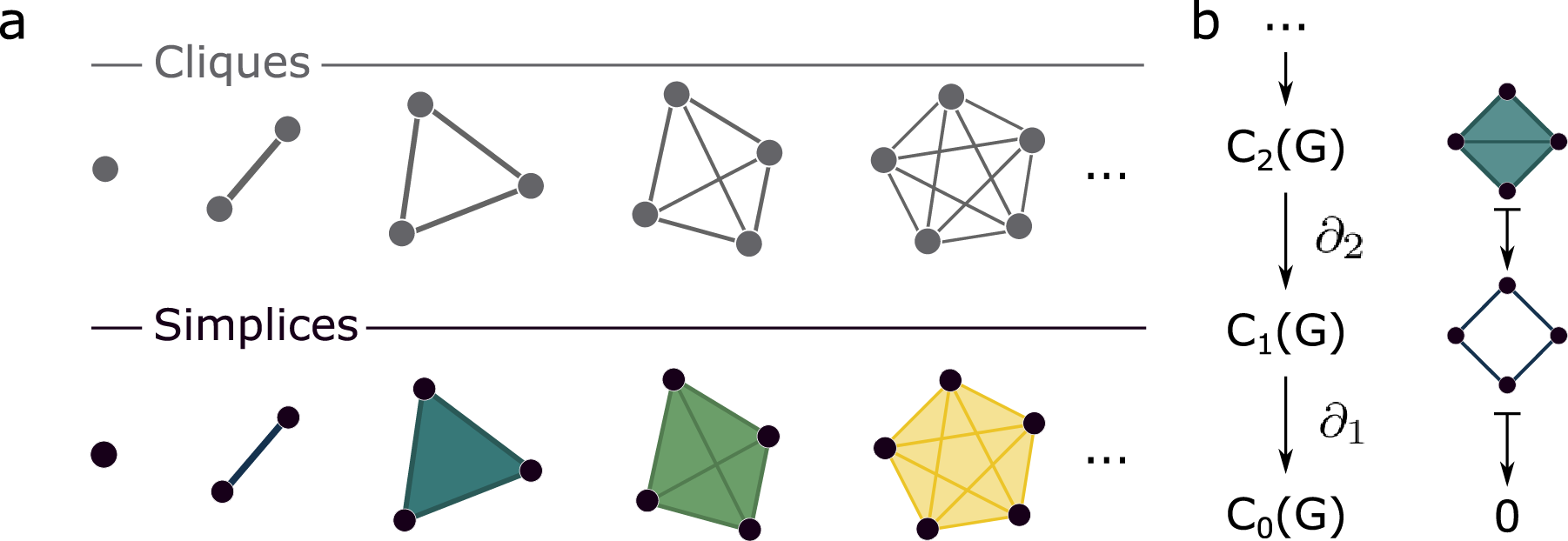}
	\caption{Simplices and boundaries. \emph{(a)} A $(k+1)$-clique is a collection of $k+1$ all-to-all connected nodes (1-clique, 2-clique, etc. shown in top). Cliques in a graph are replaced with simplices of the same number of nodes (0-simplex, 1-simplex, etc. shown at bottom). \emph{(b)} Example of an element in $C_2(G)$ sent to its boundary in $C_1(G)$, then when the boundary is again taken this is sent to 0 in $C_0(G)$.}
	\label{fig:sfig_ph1}
\end{figure}

Next we formally describe persistent homology. We begin with the task of detecting topological cavities in an unweighted graph. Given $G = (V,E)$ we translate this binary, unweighted graph into a combinatorial object called the clique complex by ``coloring in'' all cliques (all-to-all connected subgraphs) of $G$. Formally every $(k+1)$-clique, a completely connected subgraph of $G$ containing $k+1$ nodes (Fig.~\ref{fig:sfig_ph1}a, top), is replaced with a $k$-simplex (Fig.~\ref{fig:sfig_ph1}a, bottom). A $k$-simplex $\sigma = \{v_0, v_1, \dots, v_k\}$ is the convex hull of $k+1$ affinely positioned nodes. The collection of simplices created from the cliques of $G$ is called the \emph{clique complex} $X(G)$. The clique complex of $G$ is an abstract simplicial complex, meaning that we have a vertex set $V$ (the original vertex set of $G$) along with a collection $K$ of subsets of $V$ that is closed under taking subsets. So elements of $K$ are simplices, and using geometric intuition we see clearly that any subset of a simplex must also be a simplex, called a \emph{face}.  We can write $X(G) = \{X_0(G), X_1(G), \dots, X_M(G)\}$ with each $X_k(G)$ being the collection of $k$-simplices of $G$ called the $k$-skeleton. To summarize thus far we have taken our graph $G$ and translated it into the combinatorial object called the clique complex $X(G)$.

To locate topological cavities, we will need to perform algebra with elements of the clique complex. We create the \emph{chain group} $C_k(X(G))$, a vector space with basis elements $\sigma$ corresponding to $k$-simplices of $X(G)$. Then elements of $C_k(X(G))$, called $k$-chains, are linear combinations of these basis elements. Though we can certainly choose coefficients from any group (for example, $\mathbb{Z}$), for computational purposes we work in the field $\mathbb{Z}_2$. To streamline notation, we will write $C_i(G)$ to mean $C_i(X(G))$ for simplicity.\\

\noindent\emph{Boundary operator.} To locate the topological cavities, we need to first comprehend the makeup and arrangements of simplices in our simplicial complex. For example, if we only have edges, we cannot tell which closed loops are true cavities without information about the positions of higher dimensional simplices within the complex. In particular, when searching for $k$-dimensional cavities we need to know the $k$-dimensional footprints of $(k+1)$-dimensional simplices. These footprints are the boundaries of $(k+1)$-simplices that can be computed using the boundary operator $\partial_{k+1}$. The boundary operator $\partial_{k+1}: C_{k+1} \rightarrow C_k$ is defined

$$\partial_{k+1}(\sigma_{v_0, v_1, \dots, v_k}) = \sum_i (-1)^i\sigma_{v_0, v_1, \dots, v_{i-1}, v_{\hat{i}}, v_{i+1}, \dots, v_k}$$

with $v_{\hat{i}}$ omitted. Note the $(-1)^i$ records the directionality of chains, but since we work in $\mathbb{Z}_2$ we can drop this term.

The boundary operator allows us to detect cavities due to a few particularly useful properties. First, the boundary operator extends linearly, so for $a,b \in C_k$, $\partial_k(a+b) = \partial_k(a) + \partial_k(b)$. Geometrically this means that the boundary of a collection of simplices is what we would intuit: the $(k-1)$-simplices that form a ``shell'' around the $k$-chain $a+b$ (for example, see Fig.~\ref{fig:sfig_ph1}b).

Next, let us examine what happens when we take the boundary of a cycle, a closed path of simplices. Again following geometric intuition, the boundary of a cycle is the end minus the beginning, which are the same in a closed path, and therefore the boundary should be $0$. Indeed, cycles of dimension $k$ are precisely the elements in $C_k$ sent to $0$ by $\partial_k$, or $\ker(\partial_k)$. Now note that one way we could construct a cycle is to take a $(k+1)$-simplex $\sigma$ and remove the interior -- equivalently send $\sigma$ to its $k$-boundary. Then $\partial_{k+1}(\sigma)$ is a cycle, and thus $\partial_k(\partial_{k+1}(\sigma)) = 0$. If the boundary of any simplex is a cycle, then by linearity we get that the boundary of any $(k+1)$-chain is a $k$-cycle. Thus $\text{im}(\partial_{k+1}) \subseteq \ker(\partial_k)$. We call elements of $\text{im}(\partial_{k+1})$ $k$-boundaries. To summarize, we have $\ker(\partial_k)$ the $k$-cycles, $\text{im}(\partial_{k+1})$ the $k$-boundaries, and $\partial_k \circ \partial_{k+1} = 0$ (so $\text{im}(\partial_{k+1}) \subseteq \ker(\partial_k)$). \\

\begin{figure}
	\centering
	\includegraphics[width = 5in]{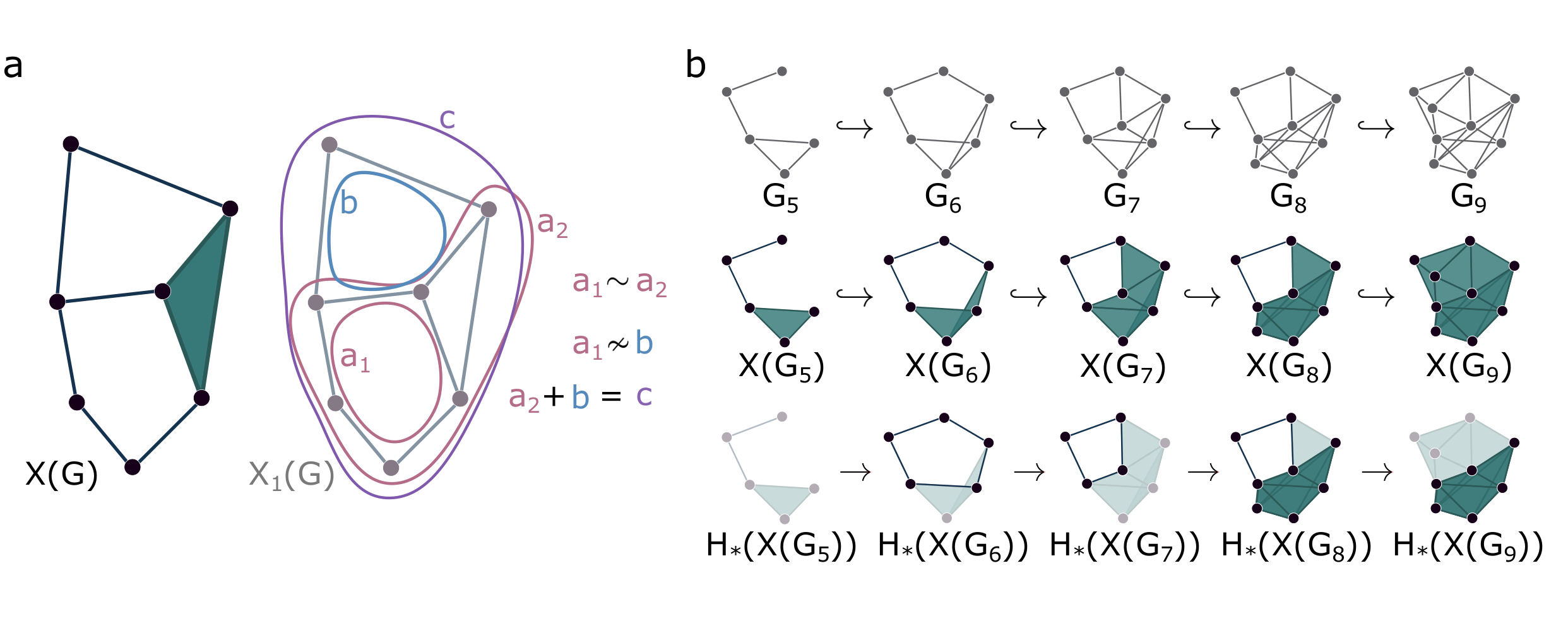}
	\caption{Equivalent cycles and filtrations. \emph{(a)} For the clique complex $X(G)$, 1-cycles $a_1$ and $a_2$ are equivalent since their difference is a boundary of a 2-simplex. However, these cycles are not equivalent to either cycle $b$ or $c$. \emph{(b)} Filtration on graphs $G_i$ (top) induces a filtration on their clique complexes $X(G_i)$ (middle) which finally induces maps between homology groups $H_*(X(G_i))$ (bottom). The minimal 1-cycle surrounding a cavity born at node 6 is highlighted as it shrinks and dies at node 9. A minimal 2-cycle born at node 8 is also highlighted.}
	\label{fig:sfig_ph2}
\end{figure}

\noindent\emph{Equivalent cycles.} We have just seen how all $k$-boundaries are necessarily $k$-cycles. But what if $\text{im}(\partial_{k+1}) \subsetneq \ker(\partial_k)$? Then there exist $k$-cycles in $\ker(\partial_k) - \text{im}(\partial_{k+1})$ that do not surround a collection of higher dimensional simplices. Thus, they must instead enclose a void of dimension $k+1$ called a $k$-cavity. Since the cavities themselves are the features of interest, we do not want to simply enumerate $\ker(\partial_k) - \text{im}(\partial_{k+1})$, but instead we need to have all cycles surrounding the same cavity count as one (to avoid grossly overcounting). If two cycles surround the same cavity (we will assume each only surrounds one cavity for the sake of this example) then their difference must be some collection of higher dimensional simplices. More precisely, if we let $\ell_1$ and $\ell_2$ denote these two $k$-cycles, then $\ell_1 - \ell_2 \in \text{im}(\partial_{k+1})$. We call these two cycles \emph{equivalent}. In fact, we say that any two cycles $a,b \in \ker(\partial_k)$ are equivalent if $a-b \in \text{im}(\partial_{k+1})$. For example, we see in Fig.~\ref{fig:sfig_ph2}a the two cycles $a_1$ and $a_2$ are equivalent because $a_2 - a_1$ is the boundary of a $2$-simplex. However, $a_1 \not\sim b$, since their difference is not a boundary of a collection of 2-simplices. We could also take the $1$-cycle $c$ which surrounds both cavities, though note that it is not equivalent to any of $a_1$, $a_2$, or $b$ but is instead the sum $a_2 + b \sim a_1+b$. The defined equivalence relation partitions $\ker(\partial_k)$ into equivalence classes $[\ell_0] = \{\ell \in Z_k | \ell_0 - \ell \in \text{im}(\partial_{k+1})\}$. Then each (non-trivial) equivalence class corresponds to a topological cavity within the simplicial complex. By abuse it is common to refer to an equivalence class of $k$-cycles as a $k$-cycle.\\

\noindent\emph{Homology.} At this point in the exposition, we have detailed the intuitions and definitions necessary to concretely define the homology groups of simplicial complexes. The homology group is simply the group formed by the equivalence classes as we defined above. Formally stated $H_k(X(G)) := \ker(\partial_k)/\text{im}(\partial_{k+1})$. Each non-trivial equivalence class corresponds to a topological cavity, so $\dim(H_k(X(G)))$ is the number of $k$-cavities within the simplicial complex $X$. The dimension of $H_k(X(G))$ is called the $k^{th}$ Betti number $\beta_k$ and the list $\{\beta_0, \beta_1, \dots, \beta_m\}$ are the \emph{Betti numbers} of $X(G)$.\\

\noindent\emph{Revisiting filtrations.} Earlier we introduced filtrations as a way to encode node-weighted networks. Consider one unit of the filtration, the map $i_k:G_k \hookrightarrow G_{k+1}$. Since every node and edge in $G_k$ maps to itself in $G_{k+1}$, we see that the map $i$ extends to clique complexes, with every simplex in $X(G_k)$ mapping to itself in $X(G_{k+1})$.  This gives us the map $i'_k: X(G_k) \hookrightarrow X(G_{k+1})$. Then, the filtration of graphs induces a filtration of clique complexes (Fig.~\ref{fig:sfig_ph2}b). Furthermore, the inclusion $X(G_k) \hookrightarrow X(G_{k+1})$ also means that we can easily map elements of the chain groups $C_*(X(G_k)) \hookrightarrow C_*(X(G_{k+1}))$, since we can take the above inclusion $i'_k$ as mapping the basis elements of $C_*(X(G_k))$ to those in $C_*(X(G_{k+1}))$. We depict these concepts in Fig.~\ref{fig:sfig_ph2}b.

Finally since we have these nice maps from one chain complex into the next, we can map cycles to cycles and consequentially the homology groups $H_*(X_k) \rightarrow H_*(X_{k+1})$ (Fig.~\ref{fig:sfig_ph2}b, bottom). This means we can not only find $k$-cavities at each filtration index, but we can \emph{follow} each $k$-cavity from the first point it exists in the filtration (called the birth), as it evolves throughout the filtration, and is killed (called the death) by simplices tessellating the cavity. Some cavities never die, so we assign them a death time of $\inf$. We call the \emph{lifetime} of a persistent cycle the $death - birth$. The birth and death can be given in terms of the edge density \cite{giusti2015clique,sizemore2016classification}, filtration index, or for this study the number of nodes added. For example, the persistent 1-cycle in Fig.~\ref{fig:sfig_ph2}b is born with the addition of node 6 and dies when node 9 is added, resulting in a lifetime $=3$.

\subsection*{Models of node-filtered order complexes}

While this exposition is motivated by early semantic learning, creating simple models with controlled properties will help us gain an intuition for possible behaviors of n-order complexes. Though we describe the following models in the main text, for conceptual organization we revisit the definitions and group by model type.

\begin{figure}[h]
	\centering
	\includegraphics[width = 4in]{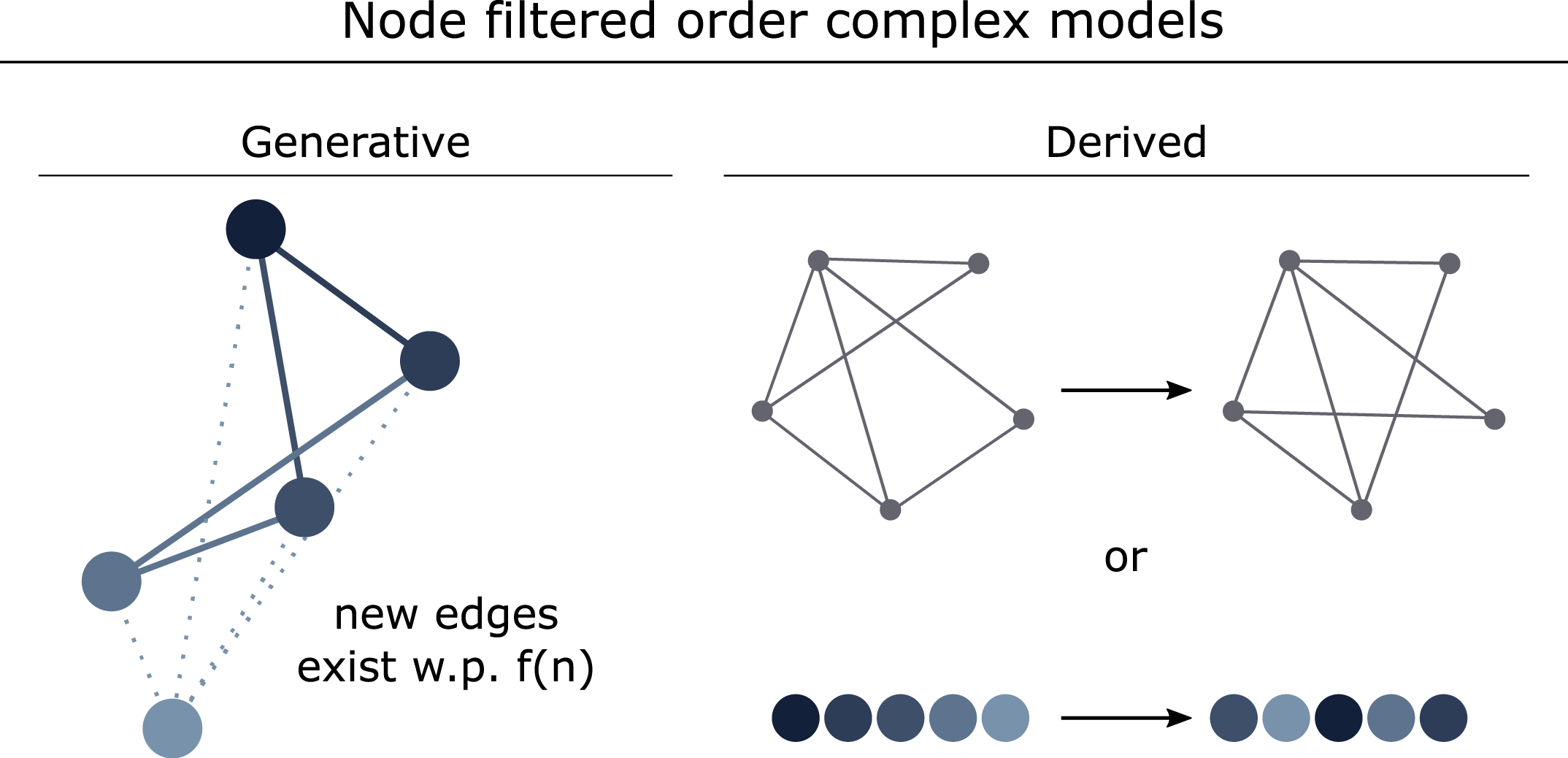}
	\caption{\textbf{Models for n-order complexes.} We design models falling into two groups: Generative and Derived. Those in the generative group are created by assigning probabilities to edge presence with each node added. Derived models can be constructed from existing complexes by altering the edges in the binary graph or by altering the order of the nodes.}
	\label{fig:sfig_models}
\end{figure}

We describe two main categories of n-order complex models: generative and derived (Fig.~\ref{fig:sfig_models}). Recall a n-order complex can be completely defined by the pair $(G,s)$ with $G$ a binary graph and $s$ an ordering of the nodes. A generative model creates the complex according to a set of rules: for n-order complexes we can generate a pair $(G,s)$ by first assuming $s = 1, 2, \dots, N$ and then constructing $G$. Derived models instead begin with either $G$ or $s$ and use this to construct the model.

We can construct simple n-order complex models using a function $p:\{1,2, \dots, N\} \rightarrow [0,1]$ such that when node $n$ is added, each edge between node $n$ and all previous nodes exist with probability $p(n)$. In the main text we include results for

\begin{equation}
p(n) = c\\,
\end{equation}
\begin{equation}
p(n) = (n/N)^d\\,
\end{equation}

called the constant probability model and proportional probability model, respectively.

We could also enforce some global architecture such as community structure on the graph. If we let $c_n$ be the community of node $n$, then we can iteratively build a binary graph $G$ at step $n$

\begin{equation}
p(n,m) = \Big\{ \begin{array}{ll}
p_{in} & c_n = c_m; \\
p_{out} & c_n \neq c_m \\
\end{array}
\Big\}
\end{equation}

with $p_{in} > p_{out}$ defining within or between module edge probabilities and $m < n$. We assign the community affiliation vector randomly with the desired number of communities and call this the modular n-order complex model.

Instead of constructing a graph with a particular global structure, it may be the case that we may have some local information such as a predetermined affinity of each node for connections. The node affinity does not change as the network grows, making this inherently different than for example the preferential attachment model \cite{barabasi1999emergence}. Then given an affinity vector $a = (a_1, a_2, \dots, a_N)$ with $a_m$ the affinity of node $m$, we can construct a n-order complex using the following rule: when the $n$th node is added, the probability of an edge forming between node $n$, $m$ is
\begin{equation}
p(n,m) = c\frac{a_m}{\max(a)} \mathrm{~.}
\end{equation}

We call this the edge affinity model. If a node with normalized affinity $=1$ is not ideal, one can multiply $\frac{a_m}{\max(a)}$ by a constant $c$ to adjust the maximum probability that any node will acquire edges after it is added.

The second class of models that we consider we call derived models, which, in contrast to generative models, alter features of an existing network and therefore require some prior knowledge of the system. We further group these into two basic types based on whether the edges of the binary graph $G$ or the node ordering $s$ changes. In the first, we maintain the original node ordering $s$ but, as an example, could randomly rewire the edges of $G$ while preserving degree distribution (similar to the configuration model \cite{bender1978asymptotic,maslov2002specificity}) which we call the randomized edges model. In the second, we maintain the original graph $G$ and reorder the nodes, either randomly (randomized node model) or perhaps based on a node property of $G$ such as degree (decreasing degree model) or topological distance from a given node (distance from $v_0$ model).

\subsection*{Additional parameters}

\begin{figure}
	\centering
	\includegraphics[width = 3in]{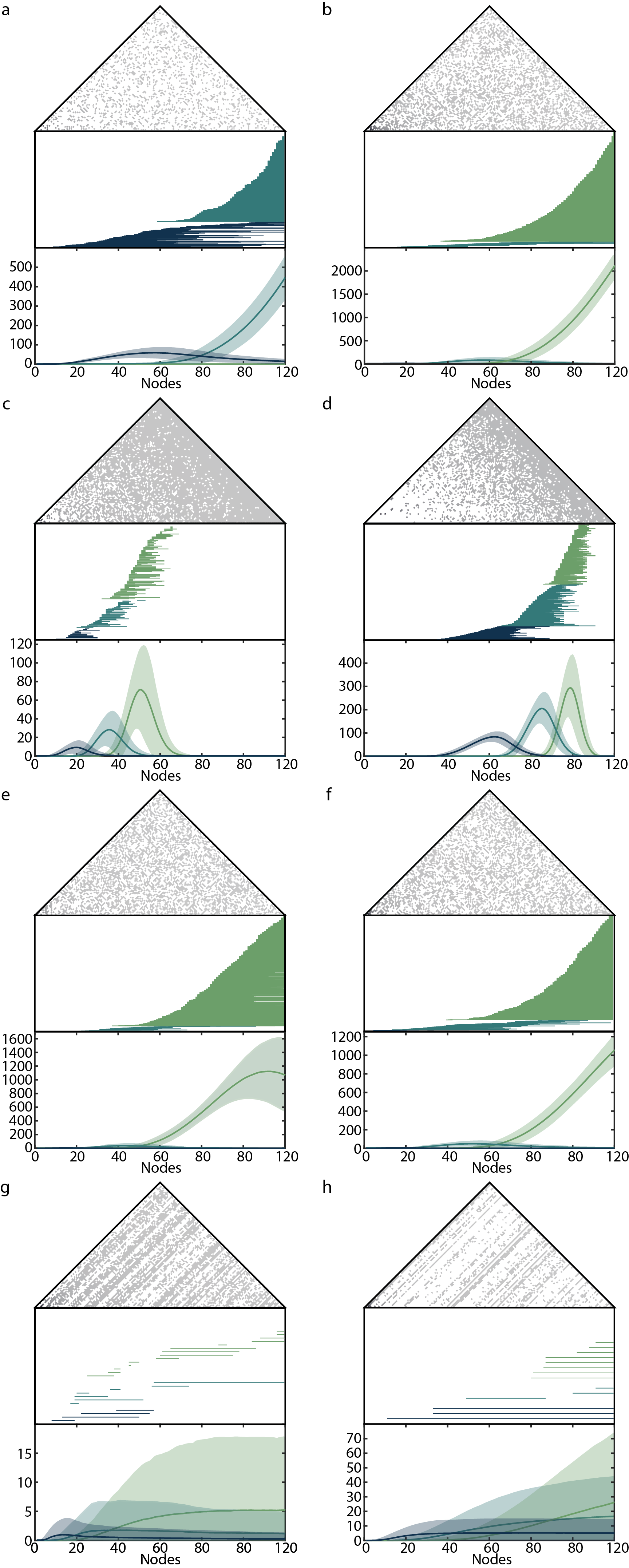}
	\caption{Persistent homology of model n-order complexes with varying parameters. Results for the constant probability model with \emph{(a)} $p=0.2$ and \emph{(b)} $p=0.4$. Persistent homology for the proportional probability model with \emph{(c)} $d = 0.5$ and \emph{(d)} $d = 2$. Results for the modular model with \emph{(e)} $p_{in} = 0.6$, $p_{out} = 0.4$ and \emph{(f)} $p_{in} = 0.8$, $p_{out} = 0.2$. Persistent homology of the edge affinity model with affinity vector a random permutation of \emph{(g)} $(1:120)$ and \emph{(h)} $(1:120)^3$.}
	\label{fig:sfig_er}
\end{figure}

Parameters for the four presented generative n-order complex models were chosen to match the edge density of the semantic feature network ($\sim 0.3$). One might ask how these parameters affect the persistent homology of the growing graphs. In Fig~\ref{fig:sfig_er} we show the persistent homology of the constant probability model with $p = 0.2$ and $p = 0.4$, the proportional probability model with $d = 0.5$ or $d =2$, the modular model with $p_{in} = 0.8$, $p_{out} = 0.2$ and $p_{in} = 0.6$, $p_{out} = 0.4$, and the edge affinity model with affinity vectors as random permutations of $(1:120)^3$ (left) and $(1:120)$ (right). We observe that the persistent homology varies considerably between outputs of the constant probability model constructed with differing parameters, suggesting that the edge density of the network plays a large role in the persistent homology of this model.

\subsection*{Further information for maternal education levels}

\begin{figure}[h]
	\centering
	\includegraphics[width = 5in]{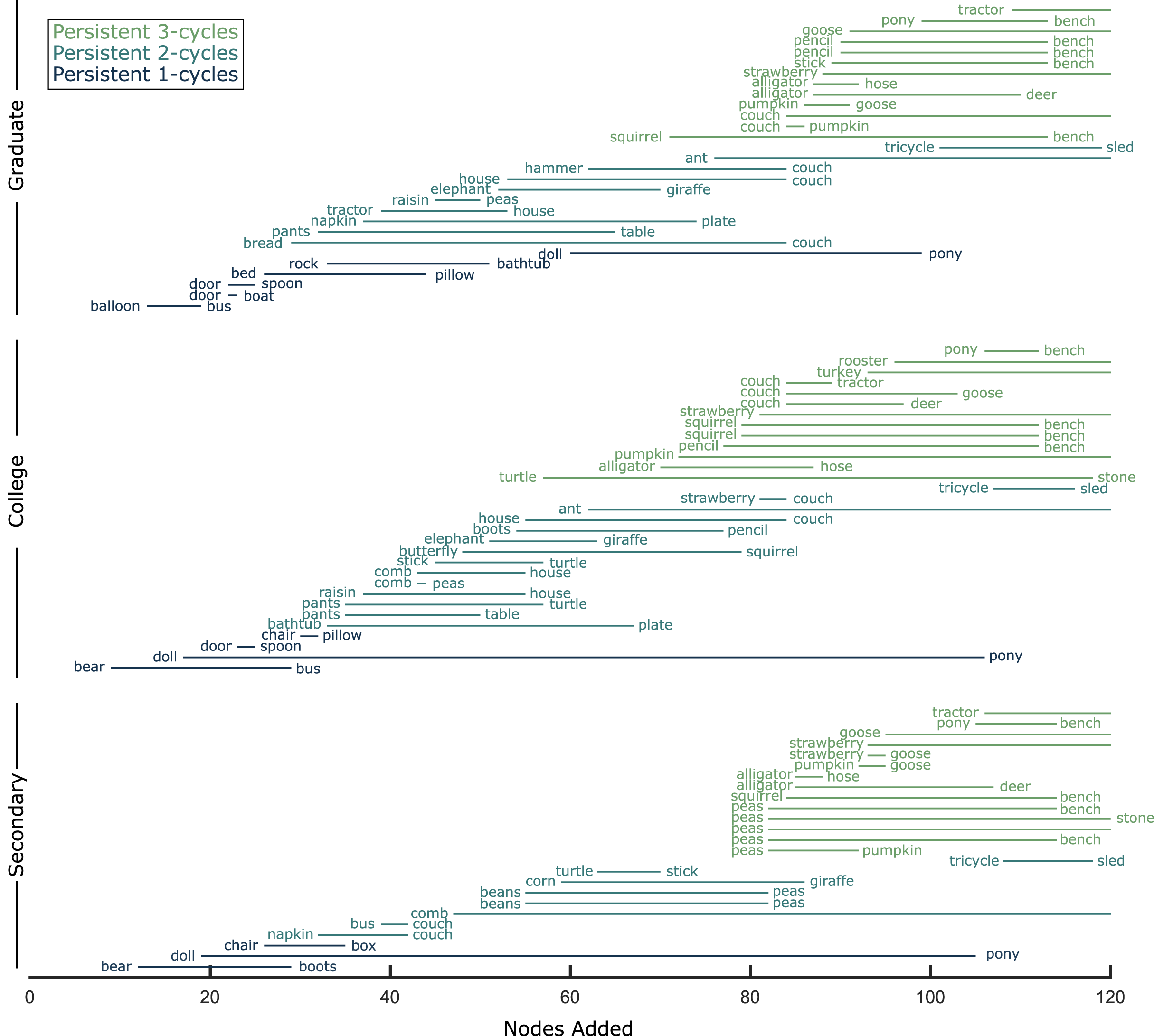}
	\caption{Barcodes for \textit{secondary}, \textit{college}, and \textit{graduate} growing networks. Words corresponding to persistent cycle birth or death nodes depicted next to the associated bar.}
	\label{fig:sfig_education1}	
\end{figure}

To supplement Fig.~\ref{fig:4}, we include in Fig.~\ref{fig:sfig_education1} the barcodes of the \textit{secondary}, \textit{college}, and \textit{graduate}
 growing semantic networks with the starting and ending words of each persistent cavity.

\subsection*{Additional persistent cycle death correlates}

\begin{figure}
	\centering
	\includegraphics[width = 5in]{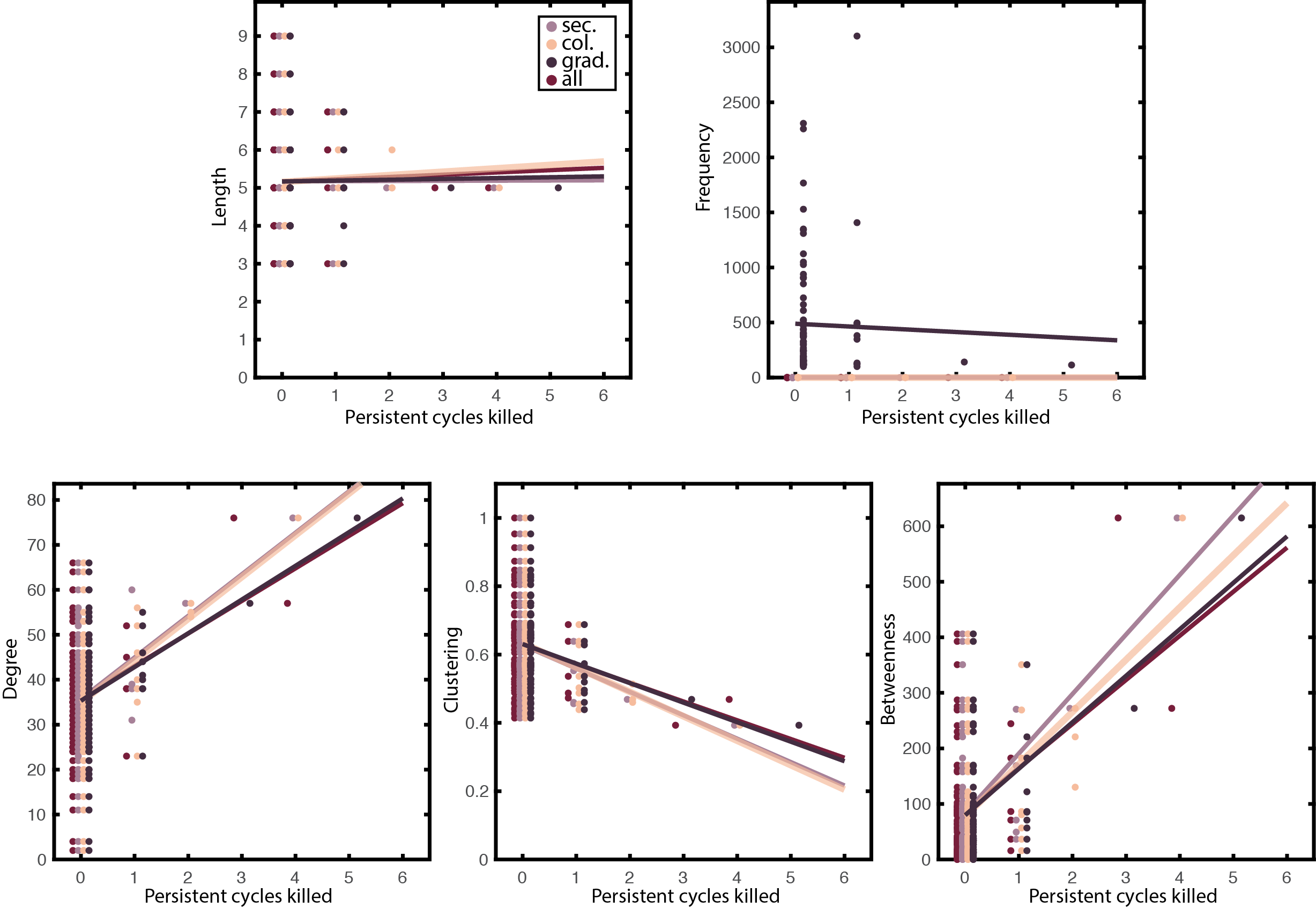}
	\caption{Additional correlates of number of persistent cycles killed.  Number of persistent cycles killed by each node against corresponding word length, frequency in child-directed speech, node degree, clustering coefficient, and betweenness centrality.}
	\label{fig:app_corr}
\end{figure}

Frequency of a word in child-directed speech is known to correlate with age of acquisition
\cite{goodman2008does}. We asked if caregiver output frequency correlates with the number of persistent cycles each node kills. We extracted frequency counts of child-directed speech from \cite{macwhinney2009childes,li2000acquisition,goodman2008does}. Of our original 120 words, 87 were found in this database so we restrict the following calculations to those 87 words. As described in the main text, we calculated the Pearson correlation coefficient between the persistent cycle death count and the node degree, clustering coefficient, betweenness centrality, word length, and word frequency for the semantic network using all children, and separately when broken down by education level (Fig.~\ref{fig:app_corr}). We observe a slightly decreasing trend of word frequency as corresponding nodes kill more persistent cycles, but this is not significant ($p > 0.3$ for each of the education levels). Trends for number of persistent cavities killed with node degree, clustering, and betweenness are similar to those seen with all nodes (Fig.~\ref{fig:6}) and remain significant ($p<0.01$) in all cases except for the clustering coefficient of the original all-included growing semantic network ($p = 0.022$). To summarize, we observe the connectivity patterns of words better determine the tendency of a word to fill in a knowledge gap than do simple lexical features.

\end{document}